\documentclass[nofootinbib,preprintnumbers,amsmath,amssymb,aps,prc, showkeys]{revtex4-1}

\usepackage{graphicx}
\usepackage{dcolumn}
\usepackage{bm}
\usepackage{hyperref}
\newcommand{\trento}{T\raisebox{-.5ex}{R}ENTo }
\newcommand{\trentonosp}{T\raisebox{-.5ex}{R}ENTo}

\begin{document}

\title{Pre-hydrodynamic evolution and its signatures in final-state heavy-ion observables}

\author{T. Nunes da Silva}
\email{t.j.nunes@ufsc.br}
\affiliation{%
 Departamento de F\'{i}sica, Centro de Ci\^{e}ncias F\'{i}sicas e Matem\'{a}ticas, Universidade Federal de Santa Catarina, Campus Universit\'{a}rio Reitor Jo\~{a}o David Ferreira Lima, Florian\'{o}polis, Brazil, Zip Code: 88040-900}%
 
\author{D. Chinellato}
 \email{daviddc@g.unicamp.br}
 \affiliation{%
  Instituto de F\'isica Gleb Wataghin, Universidade Estadual de Campinas, R. S\'ergio Buarque de Holanda, 777, Campinas, Brazil, Zip Code: 13083-859
}

\author{G. S. Denicol}
 \email{gsdenicol@id.uff.br}
\affiliation{%
  Instituto de F\'isica, Universidade Federal Fluminense,
  Av. Milton Tavares de Souza, Niter\'oi, Brazil, Zip Code: 24210-346,
}%

\author{M. Hippert}
 \email{hippert@ifi.unicamp.br}
 \affiliation{%
  Instituto de F\'isica Gleb Wataghin, Universidade Estadual de Campinas, R. S\'ergio Buarque de Holanda, 777, Campinas, Brazil, Zip Code: 13083-859
}

\author{M. Luzum}
 \email{mluzum@usp.br}
\affiliation{
 Instituto de F\'isica, Universidade de S\~ao Paulo, R. do Mat\~ao, 1371, S\~ao Paulo, Brazil, Zip Code: 05508-090
}%

\author{J. Noronha}
 \email{jn0508@illinois.edu}
\affiliation{%
  Department of Physics, University of Illinois at Urbana-Champaign,
  1110 West Green Street, Urbana, USA, Zip Code: 61801-3003
}%

\author{W. Serenone}
 \email{serenone@ifi.unicamp.br}
 \affiliation{%
  Instituto de F\'isica Gleb Wataghin, Universidade Estadual de Campinas, R. S\'ergio Buarque de Holanda, 777, Campinas, Brazil, Zip Code: 13083-859
}

\author{J. Takahashi}
 \email{jun@ifi.unicamp.br}
 \affiliation{%
  Instituto de F\'isica Gleb Wataghin, Universidade Estadual de Campinas, R. S\'ergio Buarque de Holanda, 777, Campinas, Brazil, Zip Code: 13083-859
}

\collaboration{The ExTrEMe Collaboration}

\date{\today}

\begin{abstract}
We investigate the effects of pre-hydrodynamic evolution on final-state observables in heavy-ion collisions using state-of-the art event simulations coupled to different pre-hydrodynamic scenarios, which include the recently-developed effective kinetic transport theory evolution model K{\o}MP{\o}ST. Flow observables are found to be insensitive to the details of pre-hydrodynamic evolution. The main effect we observe is in the $p_T$ spectra, particularly the mean transverse momentum. However, at least part of this effect is a consequence of the underlying conformal invariance assumption currently present in such approaches, which is known to be violated in the temperature regime probed in heavy-ion collisions. This assumption of early time conformal invariance leads to an artificially large out-of-equilibrium bulk pressure when switching from (conformal) pre-hydrodynamic evolution to hydrodynamics (using the non-conformal QCD equation of state), which in turn increases the transverse momentum. Our study indicates that a consistent treatment of pre-hydrodynamic evolution in heavy-ion collisions requires the use of non-conformal models of early time dynamics.
\end{abstract}

\keywords{heavy-ion collisions, pre-equilibrium dynamics, approach to equilibrium, collective dynamics, quark-gluon plasma, large hadron collider }

\maketitle

\section{\label{sec:level1}Introduction}
Under extreme conditions of density and temperature, quantum chromodynamics (QCD) \cite{Gross:1973id,Politzer:1973fx} predicts the existence of the quark-gluon plasma (QGP), a state of matter where quarks and gluons are not confined into hadrons. Naturally occurring examples where the QGP is formed \cite{Collins:1974ky} include the primordial universe and, possibly, the interior of ultra compact astrophysical objects such as neutron stars. The experimental program of relativistic heavy-ion collisions has been developed with the goal of producing and characterizing this extreme state of matter, shedding an important light on fundamental aspects of the strong interaction. These experiments have provided mounting evidence that, at least in collisions between large nuclei, short-lived QGP matter is formed and exhibits collective behavior \cite{Heinz:2013th}. 

Given the limitation of lattice QCD methods to problems in equilibrium \cite{Philipsen:2012nu}, the large scale dynamical evolution of the QGP formed in heavy-ion collisions has been described using relativistic viscous hydrodynamics (for a review, see \cite{Romatschke:2017ejr}). However, because the QGP cools down as it expands at relativistic speeds, it will eventually hadronize. In practice, only the final stable hadrons resulting from the decays of the zoo of exotic states formed during hadronization is detected by the experiments, i.e., the properties of the QGP must be extracted without its direct detection. Phenomenology has dealt with this by employing \textit{hybrid models} (see, for instance, \cite{Petersen:2008dd}), in which different stages of the collision event are successively modeled using different numerical models. These stages are:
\begin{itemize}
    \item \textit{initial hard scattering} between nuclei, which produces hot and dense QCD matter;
    \item \textit{hydrodynamization} during which matter approaches a fluid behavior; 
    \item \textit{hydrodynamical evolution}, during which the QGP evolves according to relativistic viscous hydrodynamics, including hadronization as the QGP cools down and hadrons are formed;
    \item \textit{interacting hadron gas evolution} as the final stage for the resulting system of hadronic resonances, during which unstable states decay.
\end{itemize}
When building a simulation chain for a hybrid model, the above steps are typically mapped into:
\begin{itemize}
    \item \textit{initial condition generation code}, which models the initial state entropy (energy) density profile resulting from the collisions between the nuclei;
    \item \textit{pre-equilibrium dynamics model}, which models the early time dynamics of the system, during which it evolves from an out-of-equilibrium state to another state where relativistic viscous hydrodynamics is assumed to hold;
    \item \textit{viscous relativistic hydrodynamics code}, which is typically the workhorse of such models, describing the dynamical evolution of the QGP and its transition into a hadronic system;
    \item \textit{particlization code}, which translates the hydrodynamical degrees of freedom at freezeout into hadrons, by sampling the hydrodynamic freezeout hypersurface;
    \item \textit{hadron cascade model}, which propagates the interacting gas of hadrons and handles resonance decays;
\end{itemize}

While these setups generally reproduce experimental data with considerable precision, there remain shortcomings in the understanding of several theoretical aspects underlying such chains of numerical codes. One of these pressing questions is to understand how the dense and hot matter formed immediately after the collision approaches fluid behavior, a process usually referred to as \textit{hydrodynamization} (for a review, see \cite{Florkowski:2017olj}). This stage of relativistic heavy-ion collisions is under active scrutiny also by experimentalists, by means of studies of collisions of the so-called \textit{small systems}, such as proton-nucleous (p-A) and proton-proton (p-p), in which hints of collective behavior have been found \cite{Khachatryan_2010, 201329}.

In this work, a state-of-the art hybrid model with different pre-equilibrium dynamical scenarios is used to investigate how the latter affect final-state observables. In particular, we show that while differential flow observables, including a principal component analysis of the two-particle correlation matrix in transverse momentum, are practically insensitive to the details of pre-equilibrum dynamics, there is a non-negligible effect on the transverse momentum spectra, which is also reflected in the integrated flow. We show evidence that at least part of this effect (and possibly most) originates from the simplifying assumption of conformal invariance in the pre-equilibrium dynamics models used in the simulation chain, which results in a large out-of-equilibrium bulk pressure at the matching with the hydrodynamical model. The resulting signature of this extra bulk pressure in final-state observables is precisely the increase of mean transverse momentum. 

In the next section we briefly review the pre-equilibrium models that are employed in this work. Section \ref{sec:setup} gives the details of our numerical setup. Our results can be found in Section \ref{sec:results}, followed by our conclusions and outlook. \emph{Definitions:} We use natural units, $\hbar=c=k_B=1$, and a mostly plus signature for the Minkowski metric. 

\section{Pre-Equilibrium Modeling}\label{sec:pre-equilibrium}
As explained above, there is a strong interconnection between the different stages of a heavy-ion collision. Therefore, improving our understanding of each of the above stages has a considerable impact in our comprehension of the global picture. In this sense, a remaining key piece of the puzzle is to understand what mechanism, if any, brings the highly out-of-equilibrium matter formed immediately after the collision to a state where a hydrodynamical description may be valid. 

Hydrodynamical studies (and hybrid models) have directly employed, as initial conditions for hydrodynamics, results from models based on the color-glass condensate framework \cite{McLerran:1993ni} such as IP-Glasma~\cite{Schenke:2012wb, Schenke:2012fw}, and parametrical models of entropy (energy) deposition such as the Glauber model~\cite{Miller:2007ri, Loizides:2014vua} and \trentonosp~\cite{Moreland:2014oya}.

More recently, effective models have been employed to bridge the gap between such models and the initial (full) energy-momentum tensor at the start of hydrodynamics. In practice, such models aim at describing the system from a very early initial time $\tau_0$, up to the time at which a hydrodynamical evolution is valid $\tau_\text{hydro}$, by evolving the resulting profiles from the preceding deposition models for $\tau_0 < \tau < \tau_\text{hydro}$.

Consider a boost invariant \cite{Bjorken:1982qr} system of on-shell non-interacting massless partons that emerge isotropically from an initial hard scattering at time $\tau_0$  \cite{Broniowski:2008qk, Liu:2015nwa}. The energy-momentum tensor of such a system at a spacetime point $(\tau, \mathbf{x})$ can then be obtained by integrating the initial number density of partons in the transverse plane, $n(\tau_0,x,y)$ over a ring of radius $c\Delta\tau = c(\tau - \tau_0)$
\begin{equation}
    T^{\mu\nu}(x,y) = \frac{1}{\tau} \int d\phi\, \hat{p}^\mu \hat{p}^\nu n(x-\Delta\tau \cos\phi, y - \Delta\tau  \sin\phi),
\end{equation}
where $\hat{p}^\mu\equiv p^\mu/p_T$ is a transverse-momentum unit vector and $\tau= \sqrt{t^2-z^2}$. This procedure effectively smooths out the energy density of the system and, in a simple model of thermalization, this free streaming dynamics is interrupted at a time $\tau_\text{fs} > \tau_0$ at which the system is assumed to suddenly attain local thermal equilibrium. Clearly, this sudden transition from a model with zero coupling to a model with finite (strong) coupling is unphysical and, thus, it may only be used as a zeroth order approximation of the pre-equilibrium dynamics.

Recently, a step forward towards a more realistic scenario was proposed in \cite{Kurkela:2018vqr, Kurkela:2018wud} where the description of the evolution of the out-of-equilibrium energy-momentum tensor during this period is done via an effective kinetic theory (EKT) of weakly coupled QCD \cite{Arnold:2002zm}. In this new framework, following general ideas from the color-glass condensate, the dynamics of the system at early times after the collision is assumed to be determined by Yang-Mills equations for the classical gluon fields. Once the gluonic fields become sufficiently dilute, at a time $\tau_\text{EKT}$, the subsequent evolution of the system becomes dominated by effective kinetic processes which ultimately drive the resulting plasma towards a state where hydrodynamics may be applicable.

The model mentioned above, called K{\o}MP{\o}ST, aims at bridging this gap between the early time dynamics and the conditions necessary for the start of hydrodynamical simulations. More specifically, between time $\tau_{\text{EKT}}$ at which kinetic processes become dominant in the evolution of the plasma and $\tau_{\text{hydro}}$ at which the plasma becomes describable by relativistic hydrodynamics, the energy-momentum tensor of the system is evolved according to a linear response formalism: the energy-momentum tensor in the causal past of a given point within the out-of-equilibrium initial condition is written as a sum between a background local average plus small perturbations:
\begin{equation}
    T^{\mu\nu} (\tau_{\text{EKT}}, \mathbf{x^\prime})= \overline{T}_\mathbf{x}^{\mu\nu}(\tau_{\text{EKT}}) + \delta T_\mathbf{x}^{\mu\nu}(\tau_{\text{EKT}}, \mathbf{x^\prime}).
\end{equation}
The linearized perturbations $\delta T_\mathbf{x}^{\mu\nu}(\tau_{\text{EKT}}, \mathbf{x^\prime})$ are propagated to later times following
\begin{equation}
    \delta T_\mathbf{x}^{\mu\nu}(\tau_{\text{EKT}}, \mathbf{x^\prime}) = \int d^2 \mathbf{x^\prime} G_{\alpha\beta}^{\mu\nu} (\mathbf{x}, \mathbf{x^\prime}, \tau_{\text{hydro}}, \tau_{\text{EKT}}) \delta T_\mathbf{x}^{\alpha\beta}(\tau_{\text{EKT}}, \mathbf{x^\prime}) \frac{\overline{T}_\mathbf{x}^{\tau\tau}(\tau_{\text{hydro}})}{T_\mathbf{x}^{\tau\tau}(\tau_{\text{EKT}})},
    \label{eqn:ektpropagation}
\end{equation}
where $G_{\alpha\beta}^{\mu\nu} (\mathbf{x}, \mathbf{x^\prime}, \tau_{\text{hydro}}, \tau_{\text{EKT}})$ are the Green's functions that propagate the perturbations from $\tau_{\text{EKT}}$ to $\tau_{\text{hydro}}$. Therefore, the evolution of the energy-momentum tensor depends on the evolution of the background $\overline{T}_\mathbf{x}^{\mu\nu}(\tau_{\text{EKT}})$ and on the response functions for linearized perturbations. 

The components of the background $T^{\mu\nu}$ are the energy density $e$ and the transverse and longitudinal pressures $P_T$ and $P_L$
\begin{equation}
\label{kompusttmunu}
    T^{\mu\nu} = \text{diag}(e, P_T, P_T, P_L).
\end{equation}
The background evolution can be parametrized in terms of the rescaled time $\tau T_\text{id.}/(\eta/s)$, where $T_\text{id.}$ represents the asymptotic ideal hydrodynamics temperature and $(\eta/s)$ is the ratio between the shear viscosity transport coefficient and the entropy density, as follows
\begin{equation}
    e = \nu_g \frac{\pi^2}{30} T_\text{id.}^4 \mathcal{E} \left[ \frac{\tau T_\text{id.}}{\eta/s}\right].
\end{equation}
In the expression above, $\nu_g =$ 16 is the number of degrees of freedom for a pure gluon plasma and $ \mathcal{E} \left[ \frac{\tau T_\text{id.}}{\eta/s}\right]$ represents a universal scaling function which interpolates between the free-streaming behavior at asymptotically early times $ \frac{\tau T_\text{id.}}{\eta/s} \ll 1$ and the conformal hydrodynamic asymptotics given by the second-order gradient expression for the energy density in a Bjorken expansion
\begin{equation}
    \frac{e(\tau)}{\nu_g \frac{\pi^2}{30}T_\text{id.}(\tau)} = 1 - \frac{8 (\eta/s)}{3 \tau T_\text{id.}} + \frac{16}{9} \left(\frac{\eta/s}{\tau T_\text{id.}}\right)^2
\end{equation}
for late times $ \frac{\tau T_\text{id.}}{\eta/s} >> 1$. Using the explicit parametrization found for $ \mathcal{E} \left[ \frac{\tau T_\text{id.}}{\eta/s}\right]$ in \cite{Kurkela:2018vqr}, the evolution of the background energy-momentum tensor is obtained by matching the point with energy density $e(\tau_\text{EKT})$ to the universal scaling curve and running the components of $T^{\mu\nu}$ along this curve up to $\tau_\text{hydro}$ (again, see \cite{Kurkela:2018vqr} for the calculation and numerical implementation details).

In order to calculate the evolution of the perturbations, the phase-space distribution function $f_{x,\mathbf{p}}$ is linearized around a spatially homogeneous background distribution $\overline{f}_\mathbf{p}$ which is anisotropic in momentum space $\mathbf{p} = (p^x, p^y, p^z)$ following \cite{Keegan:2016cpi}
\begin{equation}
    f_{x,\mathbf{p}} = \overline{f}_\mathbf{p} + \int \frac{d^2 \mathbf{k}}{(2\pi)^2} \delta f_{\mathbf{k, p}} e^{i \mathbf{k}\cdot\mathbf{x}},
\end{equation}
where $\mathbf{k}$ is the wave number of the plane wave component $ \delta f_{\mathbf{k, p}}$ in the transverse plane of the boost invariant perturbation $\delta f$. The evolution of these components can be obtained by numerically solving the Boltzmann equation for the background and perturbations 
\begin{equation}
    \left( \partial_\tau - \frac{p_z}{\tau}\partial_{p_z}\right) \overline{f}_\mathbf{p} = - \mathcal{C} [ \overline{f} ],
\end{equation}
\begin{equation}
    \left( \partial_\tau - \frac{p_z}{\tau}\partial_{p_z} + \frac{i \mathbf{p}\cdot\mathbf{k}}{p}\right) \delta f_\mathbf{k,p} = - \mathcal{C}[ \overline{f}, \delta f],
\end{equation}
using a leading-order pure-gluon QCD collision kernel $\mathcal{C}$. Once the solutions are known, the response Green's functions are constructed directly from the moments of the resulting distribution function. These are utilized for propagating the initial energy and momentum perturbations up to the hydrodynamization time. Adding those to the background, the full $T^{\mu\nu}$ at $\tau_\text{hydro}$ is finally obtained. It is important to remark that this process generates off-diagonal terms due to $G_{\alpha\beta}^{\mu\nu} (\mathbf{x}, \mathbf{x^\prime}, \tau_{\text{hydro}}, \tau_{\text{EKT}})$, which effectively account for the viscous contributions for the full $T^{\mu\nu}(\tau_{\text{hydro}}, \mathbf{x})$.

The full $T^{\mu\nu}$ can then be decomposed (using the so-called Landau frame \cite{LandauLifshitzFluids}) in terms of the usual variables used in hydrodynamical simulations of the QGP evolution, namely the energy density $e$, flow velocity $u^\mu$, and shear-stress tensor $\pi^{ \mu\nu}$  
\begin{equation}
    T^{\mu\nu} = e \left(u^\mu u^\nu +  \frac{\Delta^{\mu\nu}}{3}\right) + \pi^{\mu\nu},
    \label{eqn:Tmunu-hydroDecomposition}
\end{equation}
where $\Delta^{\mu\nu} \equiv g^{\mu\nu} + u^\mu u^\nu$. This energy-momentum tensor is then used to obtain the initial conditions for the corresponding fields in the hydrodynamical simulation. We note that the underlying conformal invariance of the kinetic theory approach leads to a zero out-of-equilibrium bulk pressure $\Pi$ contribution to the overall pressure in \eqref{eqn:Tmunu-hydroDecomposition}. This point, together with the fact that the pressure for the massless gluon gas is not the same as the corresponding QCD equilibrium pressure at the temperatures involved in heavy-ion collisions, will be very important when discussing later the subsequent matching to non-conformal relativistic viscous hydrodynamics.   

In the free streaming limit of this framework, the background energy density has a simple scaling dependence with time
\begin{equation}
    e(\tau) = \frac{e_0\tau_0}{\tau}
\end{equation}
and the linearized perturbations around this background are obtained from analytical solutions to the non-interacting Boltzmann equation in Fourier space
\begin{equation}
    \partial_\tau f_{\mathbf{k\perp, p}} + i \frac{\mathbf{p\cdot k\perp}}{\lvert \mathbf{p}\lvert} f_{\mathbf{k\perp, p}} - \frac{p^z}{\tau} \partial_{p^z} f_{\mathbf{k\perp, p}} = 0,  
\end{equation}
which can be expressed in terms of Bessel functions \cite{Kurkela:2018vqr}. These analytical results can then be used to generate analogous sets of initial conditions for hydrodynamics with a free streaming pre-hydrodynamical scenario.

The public version of the numerical implementation of K{\o}MP{\o}ST allows for evolving a given initial condition using either the full EKT evolution or its free streaming limit. In fact, in this study, we have obtained results from both evolution modes, as will be detailed in the next section.

\section{\label{sec:setup} Numerical Setup}
We have simulated collisions of Pb-Pb nuclei with center of mass energy $\sqrt{s_{NN}} = 2.76$ TeV across all centralities using a hybrid model comprised of:
\begin{itemize}
    \item \trentonosp, a parametric model for generating an initial entropy profile after the nucleus-nucleus collision \cite{Moreland:2014oya};
    \item K{\o}MP{\o}ST, the kinetic theory model mentioned in the previous section that simulates pre-equilibrium dynamics \cite{Kurkela:2018vqr, Kurkela:2018wud};
    \item MUSIC, an event-by-event relativistic second-order viscous hydrodynamics code \cite{Schenke:2010nt, Schenke:2011bn,Paquet:2015lta}; 
    \item iSS, a hadronization hypersurface sampler \cite{Shen:2014vra};
    \item UrQMD, a hadronic cascade model \cite{Bass:1998ca, Bleicher:1999xi}.
\end{itemize}
The parameters employed in the generation of the initial entropy profile in \trento were obtained from a recent Bayesian analysis by Bernhard~\cite{Bernhard:2018hnz}, except for an overall normalization factor, which was obtained by matching the resulting multiplicity of charged particles to experimental data from the ALICE Collaboration \cite{Aamodt:2010cz}.
This initial entropy density is then converted to energy density using a lattice QCD-based equation of state (details below) and used as an initial condition at time $\tau_0 = 0.2$ fm. The full energy momentum tensor is then chosen to be diagonal, and is used either as an initial condition for viscous hydrodynamics at $\tau_0$ or evolved using  K{\o}MP{\o}ST up to a later time $\tau_\text{hydro} = 1.2$ fm and inserted into hydrodynamics.

In our study, MUSIC was used to perform $2D+1$ (boost invariant) viscous hydrodynamical simulations of the hydrodynamical stage, including shear and bulk viscosities. We followed \cite{Bernhard:2018hnz} and parametrized the shear viscosity to entropy density ratio as
\begin{equation}
    (\eta/s)(T) = (\eta/s)_{\text{min}} + (\eta/s)_{\text{slope}} \cdot (T-T_c) \cdot (T/T_c)^{(\eta/s)_{\text{crv}}}
\end{equation}
while the bulk viscosity to entropy density ration is given by
\begin{equation}
    (\zeta/s)(T) = \frac{(\zeta/s)_{\text{max}}}{1 + \left( \frac{T - (\zeta/s)_{T_0}}{(\zeta/s)_{\text{width}}}\right)^2}.
\end{equation}
For the shear viscosity, $(\eta/s)_{\text{min}}$ is the minimum value at $T_c$, $(\eta/s)_{\text{slope}}$ is a slope above $T_c$, and $(\eta/s)_{\text{crv}}$ is a curvature parameter. The expression for the bulk viscosity is a Cauchy distribution, which parametrizes a symmetric peak with three free parameters: a maximum value, its width, and center. The numerical values employed for these parameters were also obtained from the latest Bayesian analysis of Ref.~\cite{Bernhard:2018hnz}.

The equilibrium equation of state we used is the parametrization constructed by Huovinen and Petreczky known as \textit{s95p-v1.2 }\cite{Huovinen:2009yb}. For temperatures below 184 MeV, this equation of state is based on a hadron resonance gas. Above this temperature, it employs lattice results by Bazavov \textit{et al.}~\cite{Bazavov:2009zn}. This version of the equation of state contains the same particle species as UrQMD. We leave the implementation of a more realistic version of the QCD equation of state, e.g the one used in \cite{Alba:2017hhe,Alba:2017mqu}, to future work.  

The transition from the hydrodynamical degrees of freedom to the hadron gas is made via the Cooper-Frye formalism~\cite{PhysRevD.10.186} including viscous corrections. The freezeout hypersurfaces from hydrodynamics have been repeatedly sampled until $5\times 10^5$ particles were acquired per unity rapidity. The final configurations of stable particles were then stored in a ROOT-based C++ class, and finally used for data analysis and the calculation of observables. 

In order to disentangle the effects of pre-equilibrium dynamics over the observables of interest, three main scenarios have been studied. In all three cases, results from the \trento model are regarded as an initial entropy density profiles at time $\tau = 0.2$. The scenarios differ in the following manner:
\begin{itemize}
    \item In \textbf{Scenario A}, the initial entropy density profile generated by \trento is converted to energy density and utilized directly as the initial condition for hydrodynamical evolution, starting at time $\tau_0 = 0.2$ fm.  The transverse velocity, initial shear-stress tensor, and bulk pressure are chosen to vanish;
    \item In \textbf{Scenario B}, the same initial energy profile from Scenario A is used.  The $T^{\mu\nu}$ is determined according to Eq.~\eqref{kompusttmunu} with $P_L = 0$ (and, consequently, $P_T = e/2$), and then evolved using the free streaming limit of K{\o}MP{\o}ST from $\tau_0 = 0.2$ fm until $\tau_\text{hydro} = 1.2$ fm, at which point the hydrodynamical evolution is started;
    \item \textbf{Scenario C}, is analogous to Scenario B, but now the the initial profile is propagated from $\tau_0 = 0.2$ fm until $\tau_\text{hydro} = 1.2$ fm using the K{\o}MP{\o}ST effective kinetic theory model~\cite{Kurkela:2018vqr, Kurkela:2018wud} with $\eta/s = 0.16$, after which hydrodynamic evolution is initiated.
\end{itemize}
We have employed the same initial \trento profiles for the three scenarios above.  The only difference is a rescaling of the overall normalisation factor in the model.  In current models for initial conditions the overall normalization factor is an unknown free parameter that is chosen so that the correct final multiplicity is obtained.
We follow this procedure and adjust the energy normalization so that all the scenarios yield a charged particle multiplicity at central events consistent with experimental data.   Specifically, the initial energy from Scenario A is multiplied by a factor 0.86 for Scenario B and 1.09 for Scenario C.  That is, the free streaming evolution tends to increase the final particle multiplicity compared to hydrodynamic evolution, while the EKT evolution tends to decrease it.  

\begin{figure}[!ht]
  \includegraphics[width=.5\linewidth]{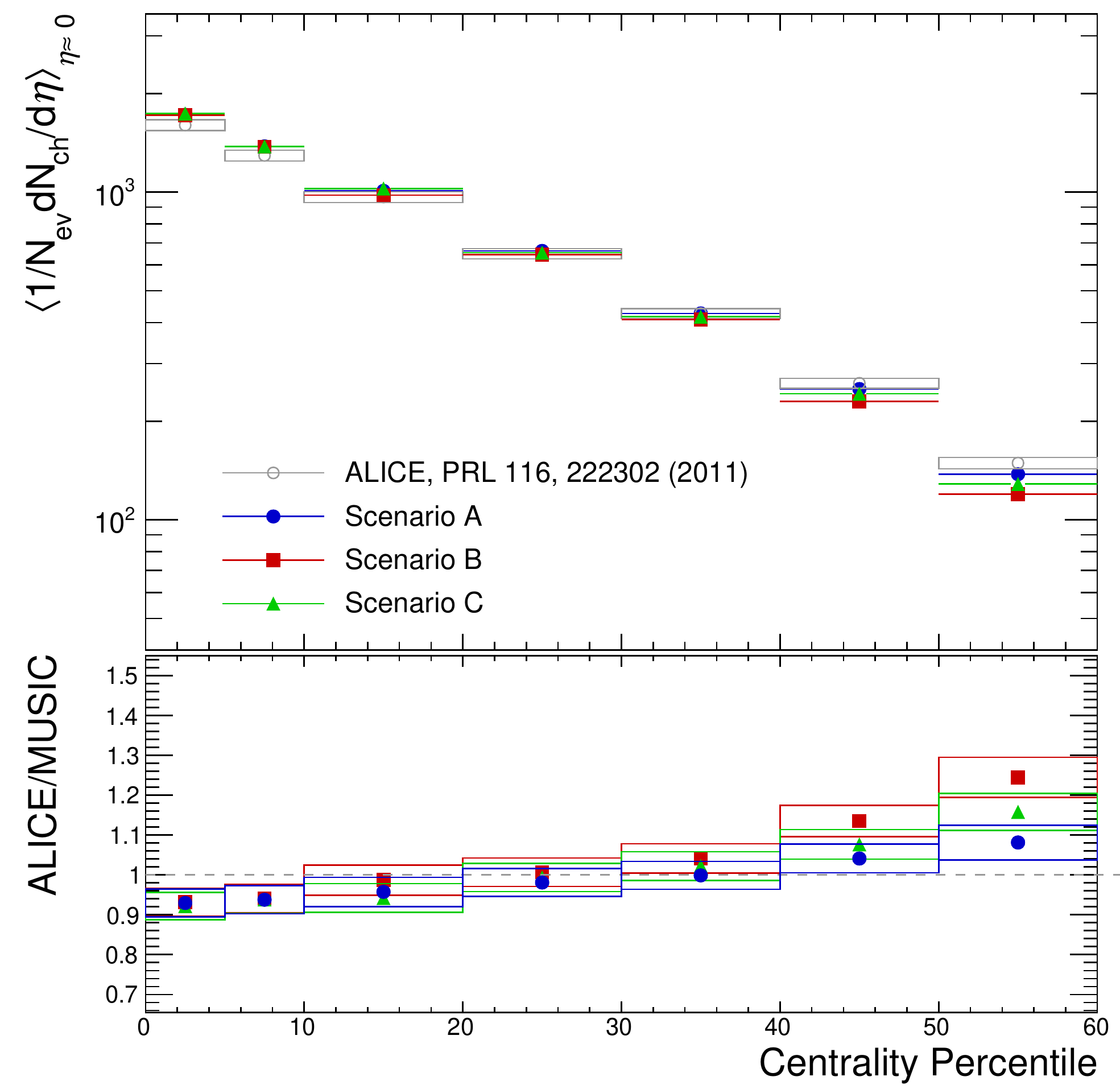}
\caption{Final multiplicity of charged particles as a function of centrality for all of the pre-hydrodynamical scenarios under consideration (top panel). We also show the ratio between the experimental results and results from our simulations (bottom panel).}
  \label{fig:mult}
\end{figure}

For each of these cases, the resulting multiplicities of final charged particles are shown as a function of centrality, in Fig.\ \ref{fig:mult}, where they are compared to experimental results from the ALICE collaboration~\cite{Aamodt:2010cz}.
These results were also used as a consistency check for the model.

\section{Results}\label{sec:results}
\subsection{Flow analysis}
\begin{figure}[ht]
  \includegraphics[width=.475\linewidth]{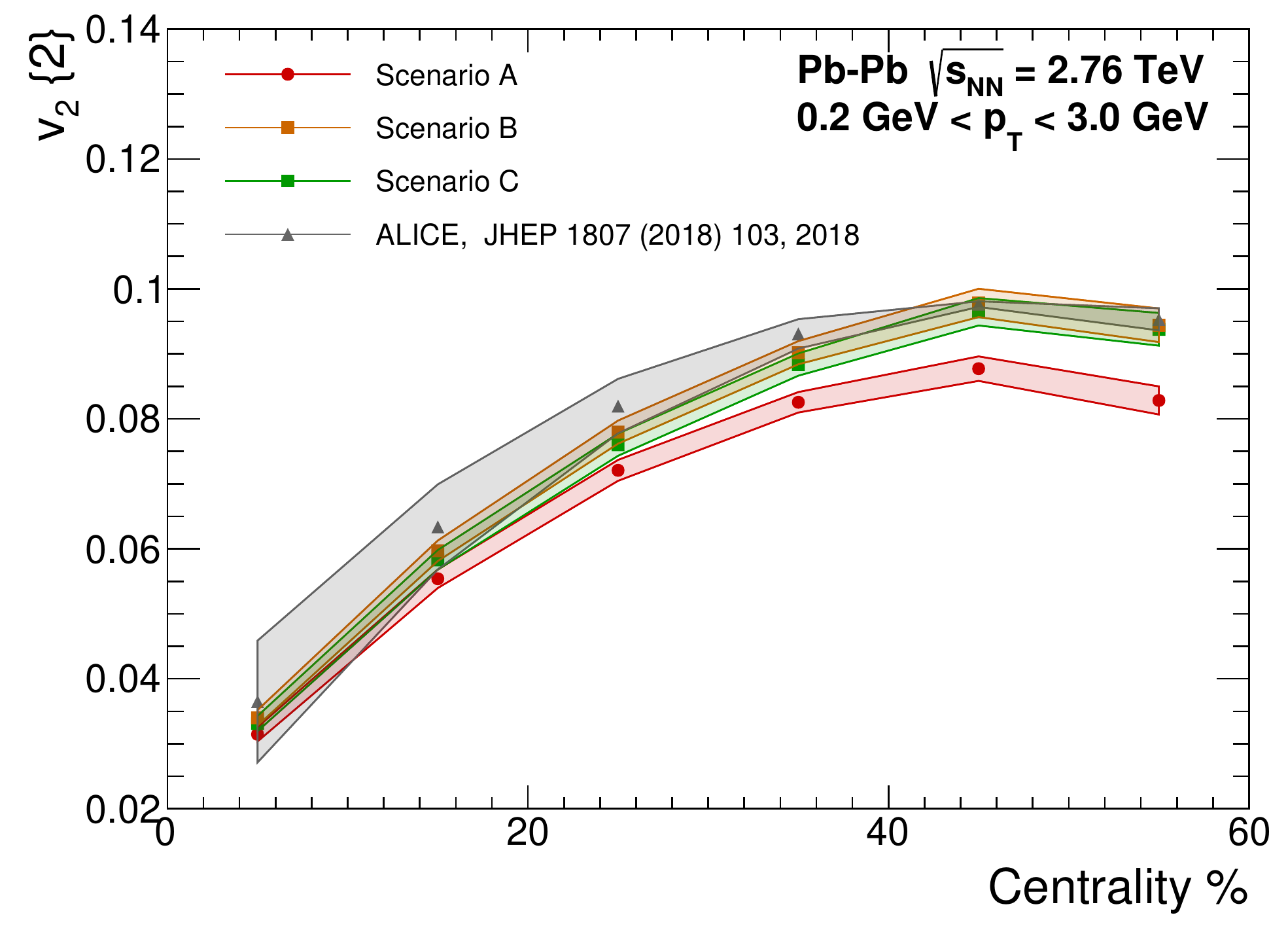}
  \includegraphics[width=.475\linewidth]{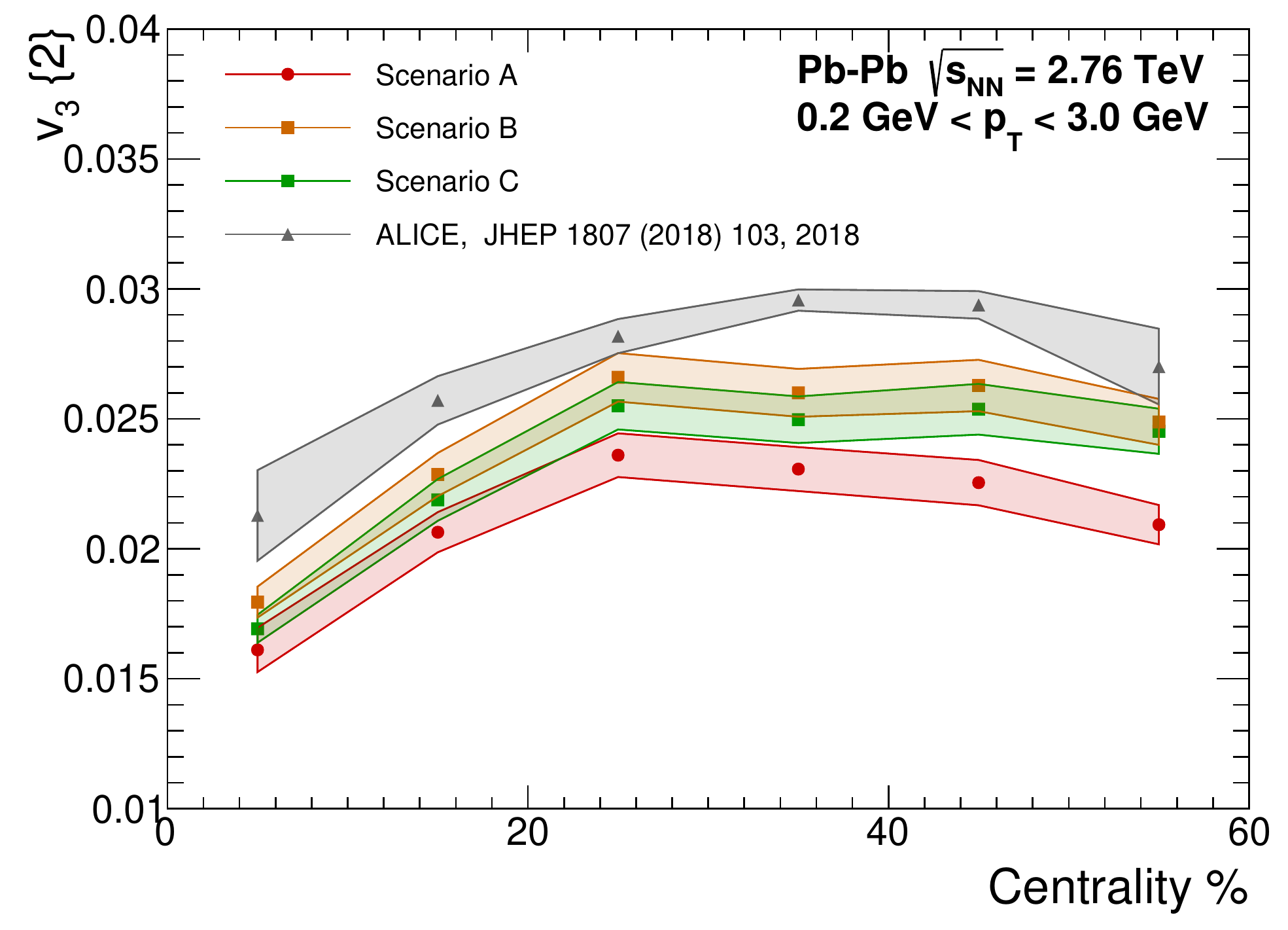}
  \caption{Anisotropic flow coefficients from two particle correlations $v_2\{2\}$ (left) and $v_3\{2\}$ (right) for the three scenarios plotted as a function of event centrality.}
  \label{fig:anisotropic-integrated}
\end{figure}

We start by analyzing the usual anisotropic flow observables $v_n$, which are the coefficients of the Fourier expansion of the probability distribution of finding a particle at rapidity $y$ with transverse momentum $p_T$ in the azimutal angle $\phi$
\begin{equation}
E\frac{dN}{d^3 p} \equiv \frac{1}{2\pi} \frac{\sqrt{m^2 + p_T^2 \cosh{(\eta)^2}}}{p_T \cosh{(\eta)}} \frac{dN}{p_T dp_T d\eta} \left[ 1 + \sum_{n=1}^\infty v_n(p_T, y) \cos{n(\phi - \Psi_n)} \right].
\end{equation}
We have extracted the anisotropic flow coefficients from the simulated events following the Q-cumulants formalism, which uses multiparticle azimuthal correlation to avoid dealing with measurements of event plane angles. Explicitly, the flow coefficients can be related to the two-particle correlation function for a given centrality through the relation \cite{Borghini:2001vi, Bilandzic:2010jr}
\begin{equation}
     v_n\{2\} = \sqrt{\langle v_n \rangle^2} = \sqrt {\langle\langle e^{in(\phi_1 - \phi_2)}\rangle\rangle}.
\end{equation}
We have followed the usual computational strategy and built the so-called Q-vector \cite{Bilandzic:2010jr}
\begin{equation}
    Q_n = \sum_{i=1}^M e^{in\phi_i},
\end{equation}
so that the $v_n$ coefficients extracted from two particle correlations is given by
\begin{equation}
    v_n\{2\} = \sqrt{ \left\langle \frac{\lvert Q_n \lvert^2 - M}{M(M-1)}\right\rangle}.
\end{equation}
Results for the integrated $v_n\{2\}$ in the transverse momentum interval $0.2 < p_t < 3.0$ GeV for the three scenarios are shown in Fig.~\ref{fig:anisotropic-integrated} for $n=2,3$. The results are compared to experimental data from the ALICE  collaboration~\cite{Acharya:2018lmh}. The presence of pre-equilibrium dynamics results in an increase of the integrated $v_n$ coefficients. As a result, the free streaming and the EKT scenarios exhibit a better agreement to experimental data. This is to be expected, since the Bayesian analysis performed to estimate the parameters used in this work was performed within a model that includes a period of free streaming evolution.  

The origin of such increment in anisotropic flow likely stems from a change in the transverse momentum dependence of these observables. In order to investigate this possibility and, more generally, to characterize flow fluctuations and momentum dependence of two-particle correlations, we have performed a Principal Component Analysis (PCA) of the two-particle covariance matrix in transverse momentum $V_{n\Delta}$ \cite{Bhalerao:2014mua, Mazeliauskas:2015efa, Mazeliauskas:2015vea}. 

This matrix can be written both in terms of its eigenvalues $\lambda_n^{(\alpha)}$ and normalized eigenvectors $\psi_n^{(\alpha)}(\mathbf{p})$, according to the spectral theorem, and in terms of its principal components $ V_n^{(\alpha)}$
\begin{equation}
    V_{n\Delta}(\mathbf{p_1},\mathbf{p_2}) \equiv \langle V_n^*(\mathbf{p_1}) V_n(\mathbf{p_2}) \rangle =\sum_{\alpha=1}^\infty \lambda_n^{(\alpha)} \psi_n^{(\alpha)}(\mathbf{p}_a) \psi_n^{(\alpha)}(\mathbf{p}_b) \\
    = \sum_{\alpha=1}^\infty V_n^{(\alpha)}(\mathbf{p}_a) V_n^{(\alpha)}(\mathbf{p}_b). 
\end{equation}
The principal components are then defined in terms of  $\lambda_n^{(\alpha)}$ and  $\psi_n^{(\alpha)}(\mathbf{p})$ as \cite{Bhalerao:2014mua}
\begin{equation}
    V_n^{(\alpha)} \equiv \sqrt{\lambda^{(\alpha)}} \psi_n^{(\alpha)} (\mathbf{p}).
\end{equation}
While the leading principal component proxies the usual $p_T$-differential flow coefficient $v_2$, the subleading modes characterize flow correlations between two different transverse momenta \cite{Bhalerao:2014mua}.

In a recent work by the ExTrEme collaboration in Ref.\ \cite{Hippert:2019swu}, it was shown that the formalism proposed by Bhalerao \textit{et al.}~\cite{Bhalerao:2014mua}, in which the covariance matrix is written as
\begin{equation}
    V_{n\Delta}^N (\mathbf{p}_a, \mathbf{p}_b ) \equiv \frac{1}{(2\pi\Delta p_T \Delta y)^2} \left\langle \sum_{a\neq b} e^{-in(\phi_a - \phi_b)} \right\rangle
\end{equation}
suffers from contamination of the subleading principal components due to multiplicity fluctuations for $n>0$. In the same work, an alternative prescription was proposed for performing the PCA analysis that removes such contaminations and makes it possible to isolate novel fluctuation sources, by diagonalizing, instead, the covariance matrix \cite{Hippert:2019swu} 
\begin{equation}
    V_{n\Delta}^R (\mathbf{p}_a, \mathbf{p}_b ) \equiv\frac{\left \langle \sum_{a\neq b} e^{-in(\phi_a - \phi_b)} \right\rangle }{\langle N_{\text{pairs}}(\mathbf{p}_a, \mathbf{p}_b)\rangle}.
\end{equation}

We have performed such a PCA analysis for all the three scenarios simulated with our model for $n=0$ following the proposal by Bhalerao et al. \cite{Bhalerao:2014mua} and for $n=2$ and $n=3$ following the ExTrEme prescription (the $n=0$ PCA analysis measures precisely multiplicity fluctuations, which are removed in the latter prescription).\footnote{A PCA analysis of the events generated with our model following the Bhalerao et al. prescription for $n=2$ and $n=3$ has been presented in a previous work \cite{NunesdaSilva:2018viu}.}
\begin{figure}
  \includegraphics[width=.475\linewidth]{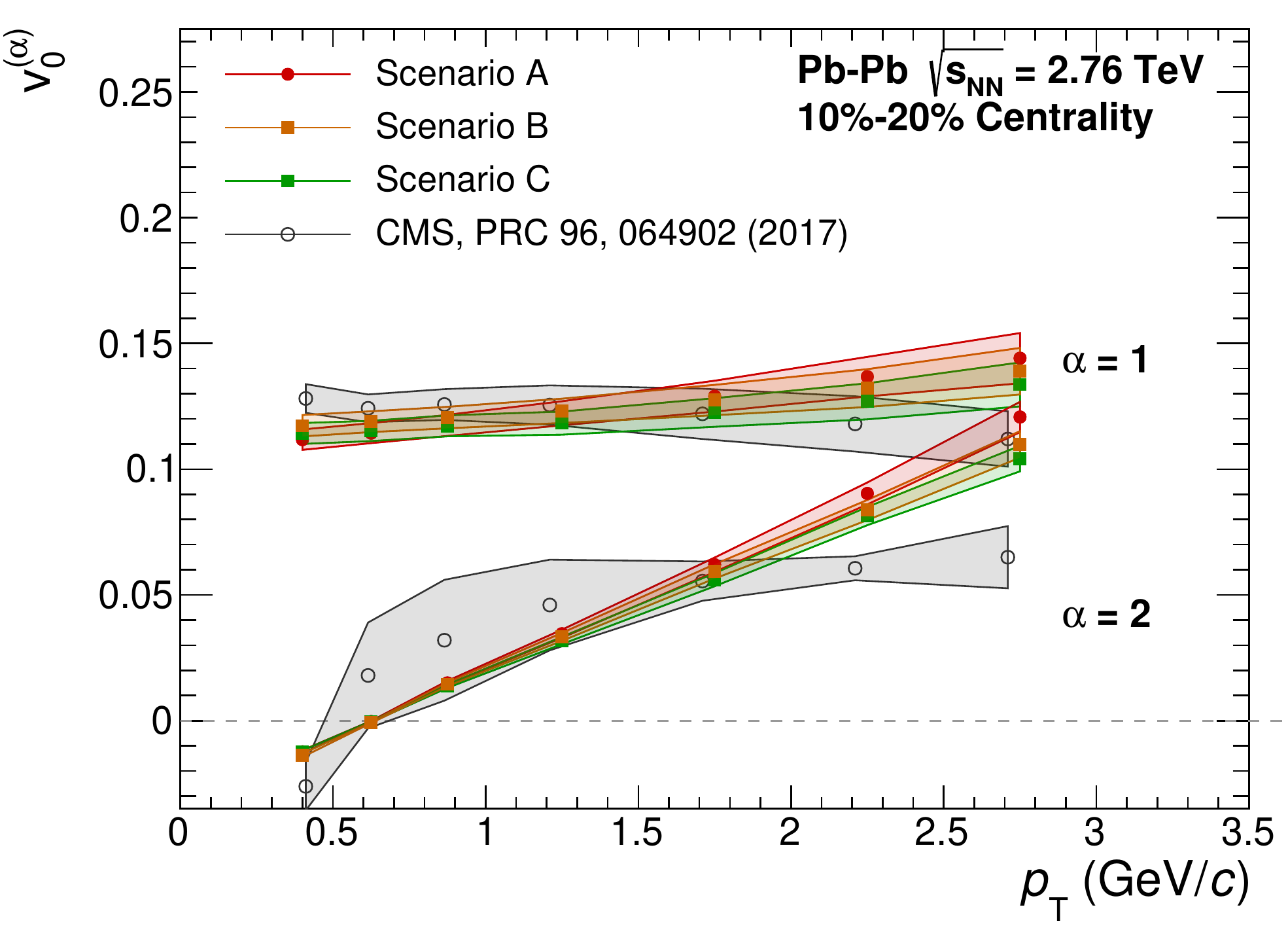}
  \includegraphics[width=.475\linewidth]{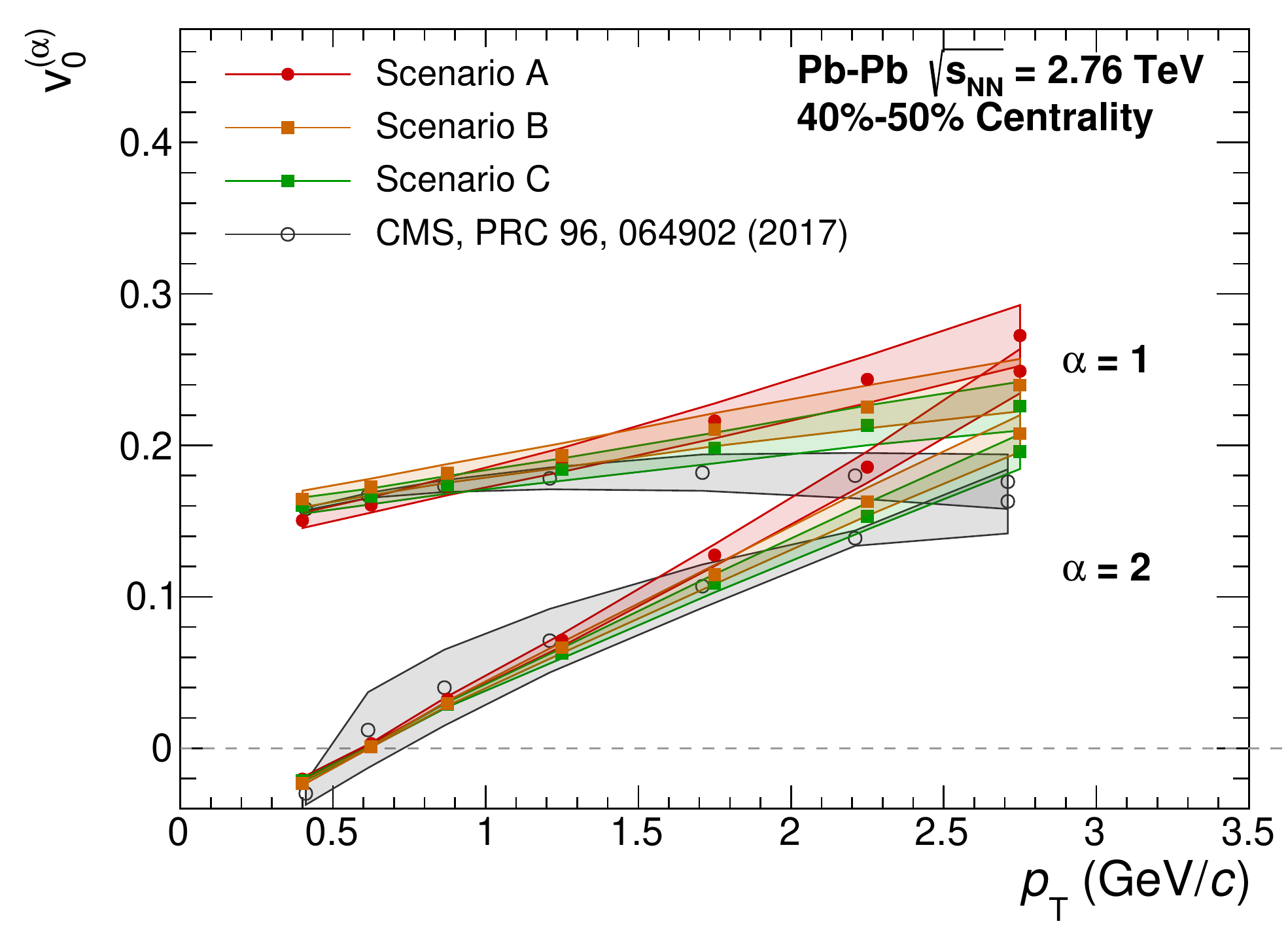}
  \caption{First two principal components of the two-particle correlation matrix according to the prescription by Bhalerao et al. \cite{Bhalerao:2014mua}, for the harmonic $n=0$ in the 10\%-20\% centrality class (left) and in the 40\%-50\% centrality classes (right) plotted as a function of the transverse momentum $p_T$.}
  \label{fig:pca0}
\end{figure}
\begin{figure}
  \includegraphics[width=.475\linewidth]{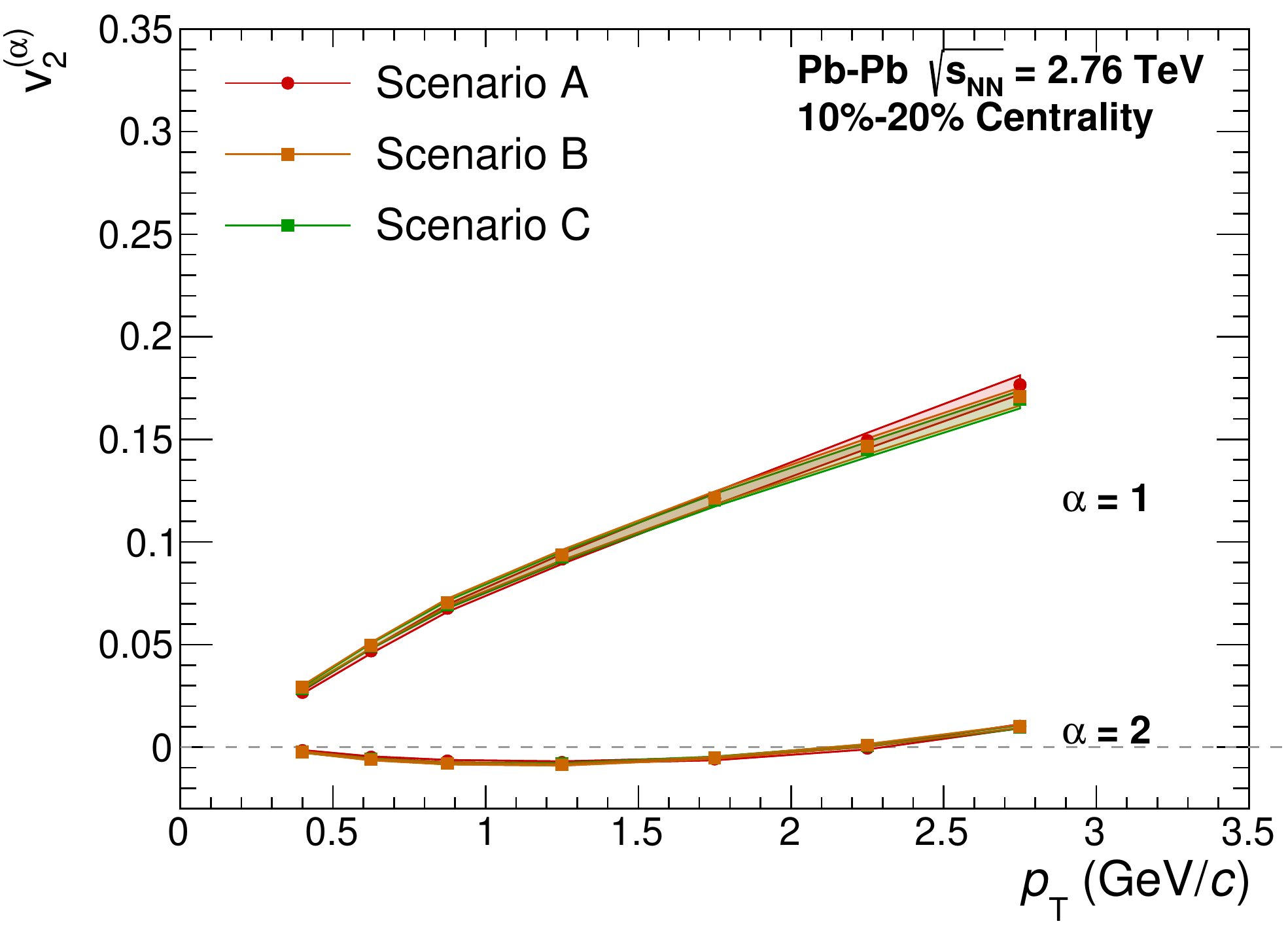}
  \includegraphics[width=.475\linewidth]{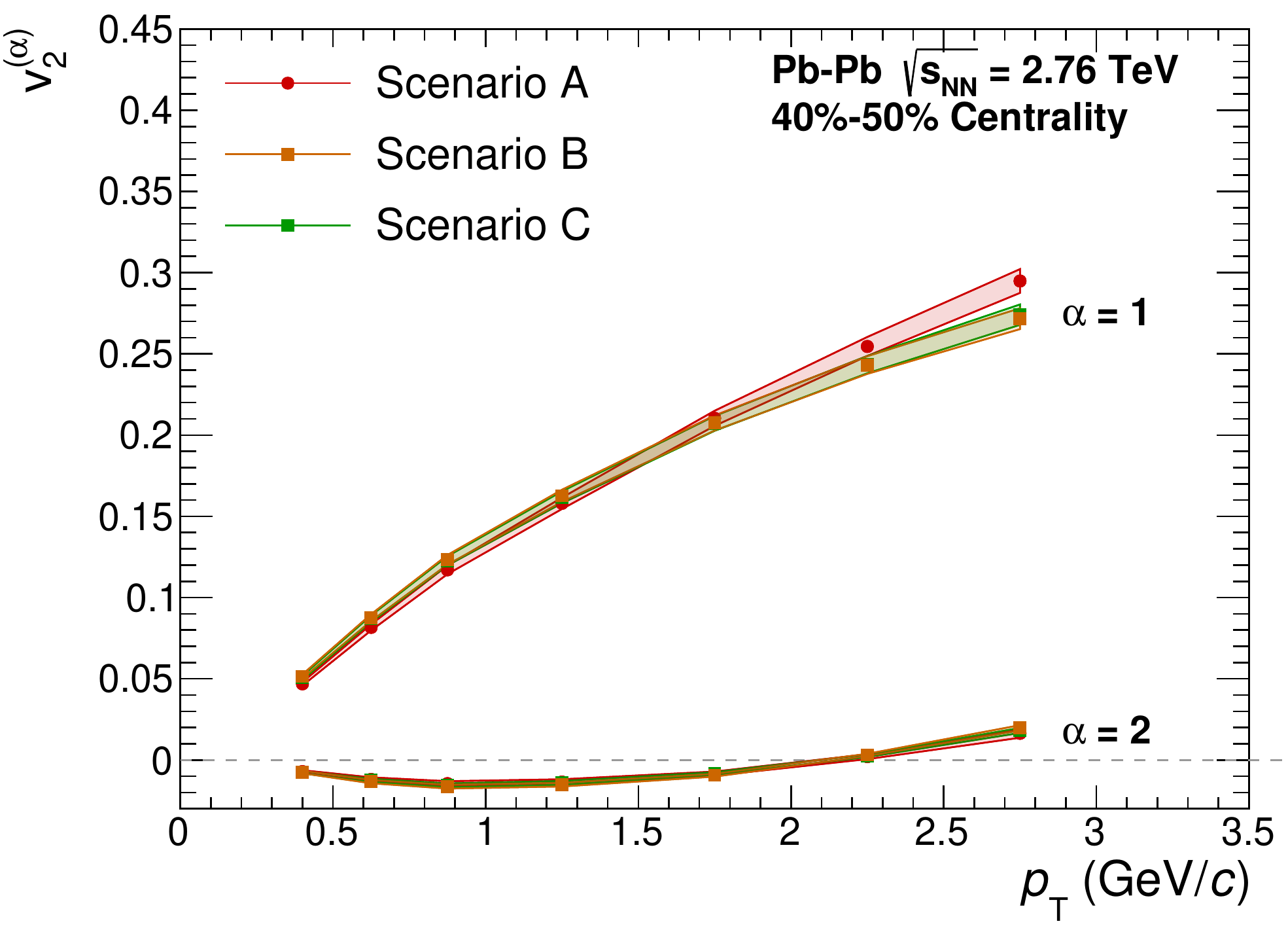}
  \caption{First two principal components of the two-particle correlation matrix according to the ExTrEMe prescription \cite{Hippert:2019swu}, for the harmonic $n=2$ in the 10\%-20\% centrality class (left) and in the 40\%-50\% centrality classes (right) plotted as a function of the transverse momentum $p_T$.}
  \label{fig:pca2}
\end{figure}
\begin{figure}
  \includegraphics[width=.475\linewidth]{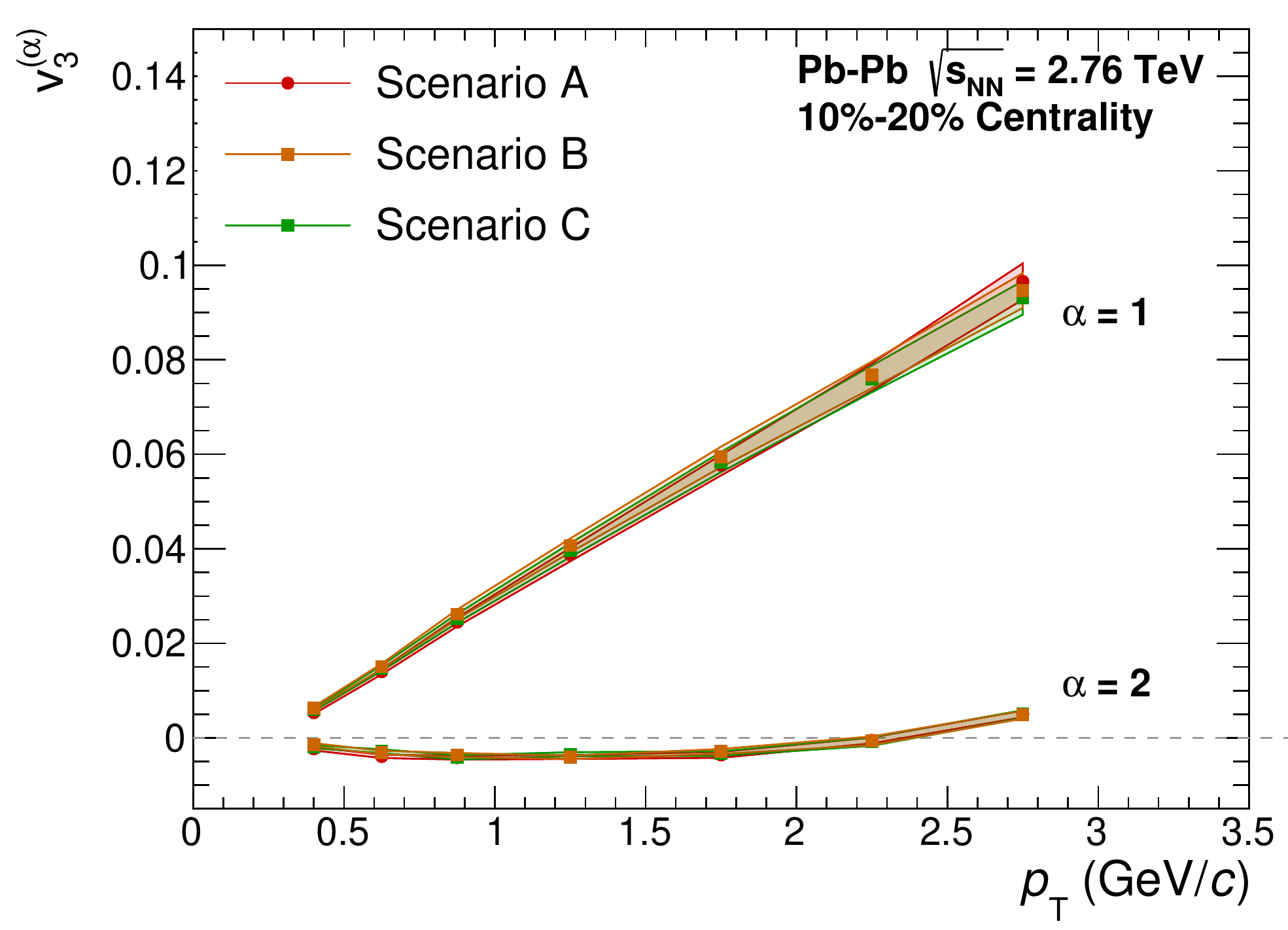}
  \includegraphics[width=.475\linewidth]{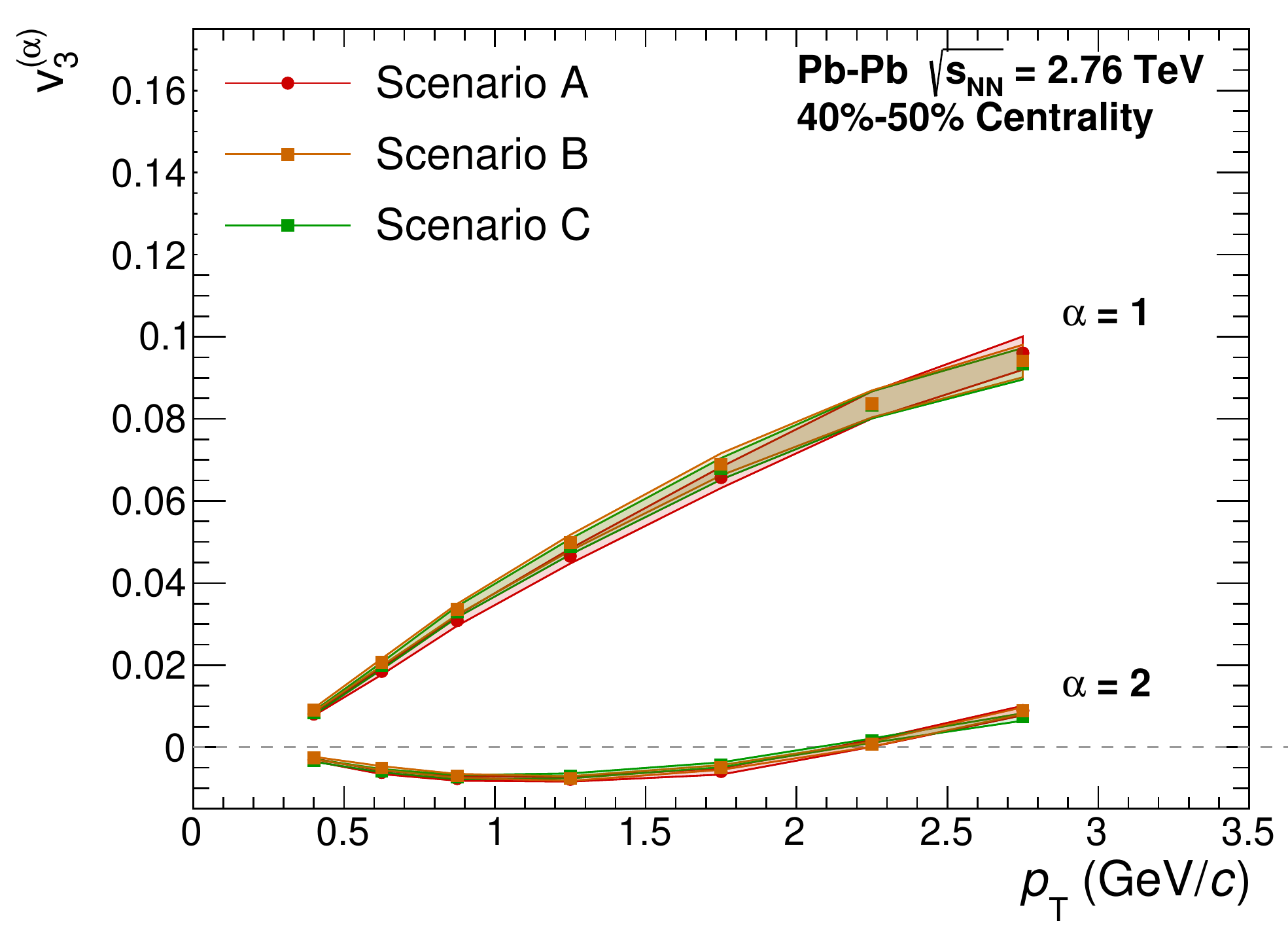}
  \caption{First two principal components of the two-particle correlation matrix according to the ExTrEMe prescription \cite{Hippert:2019swu}, for the harmonic $n=3$ in the 10\%-20\% centrality class (left) and in the 40\%-50\% centrality classes (right) plotted as a function of the transverse momentum $p_T$.}
  \label{fig:pca3}
\end{figure}

It is clear from the plots that the measured PCA components are insensitive to the type of pre-equilibrium dynamics utilized in the simulations. While this fact makes these observables particularly useful in isolating effects from the hydrodynamical evolution from effects of early stage dynamics, it also means that the source of the extra anisotropy in the integrated observables introduced by our models of pre-equilibrium dynamics does not stem from a possible change of the $p_T$ dependence of the flow.  This indicates that another effect must be taking place in the transverse momentum spectra, which would be reflected in the integrated observables.

\begin{figure}[ht]
  \includegraphics[width=.32\linewidth]{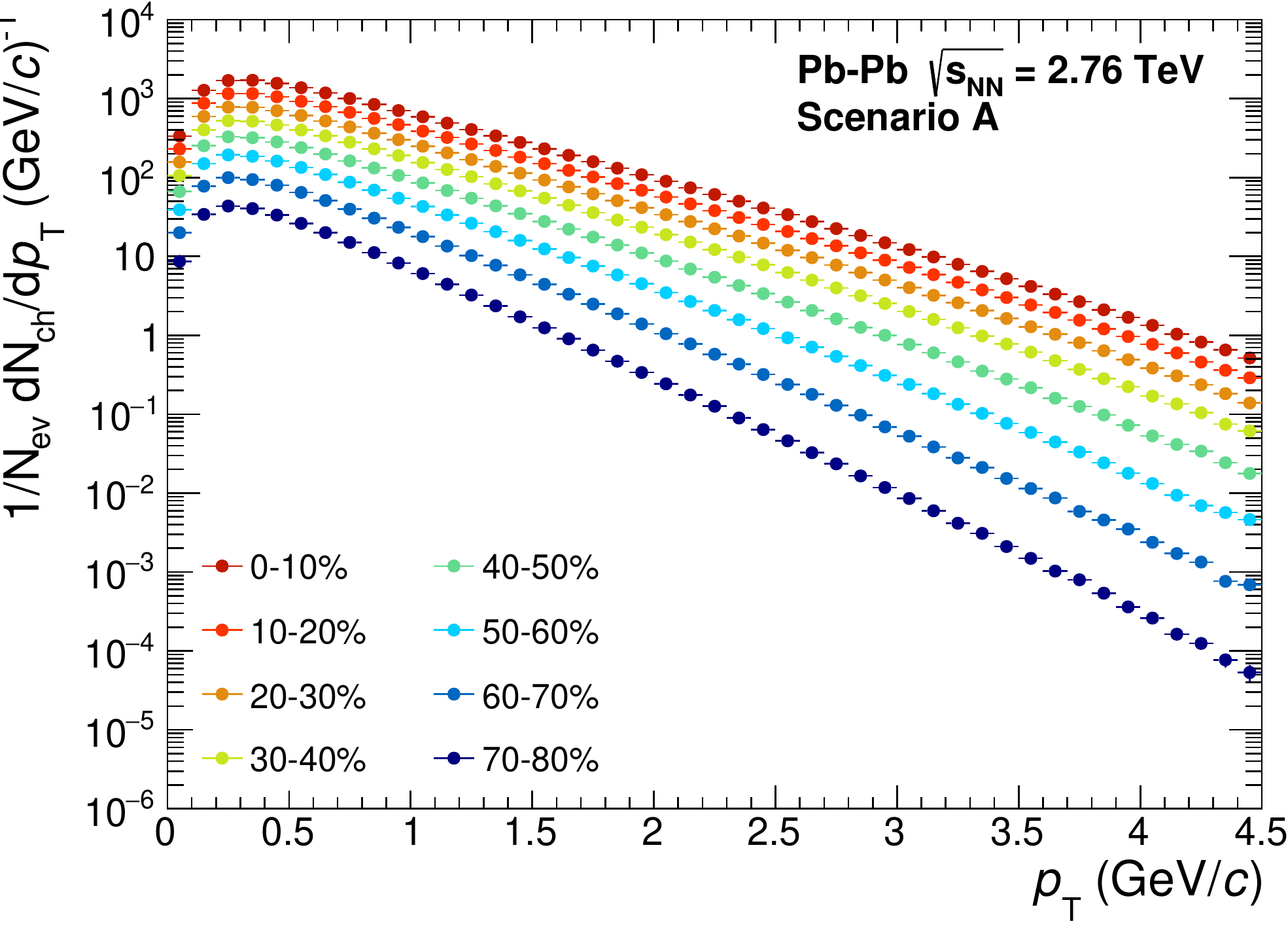}
  \includegraphics[width=.32\linewidth]{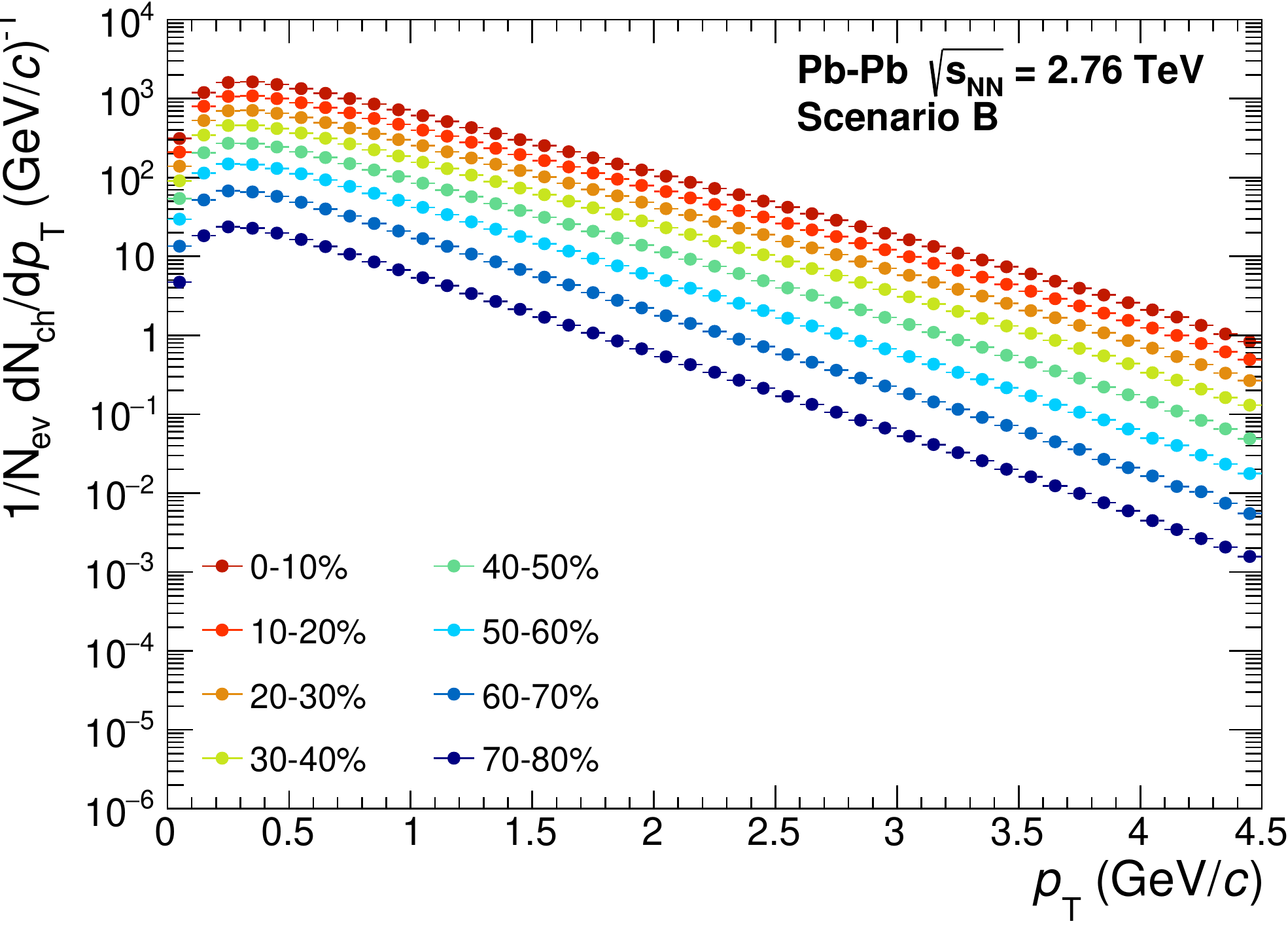}
  \includegraphics[width=.32\linewidth]{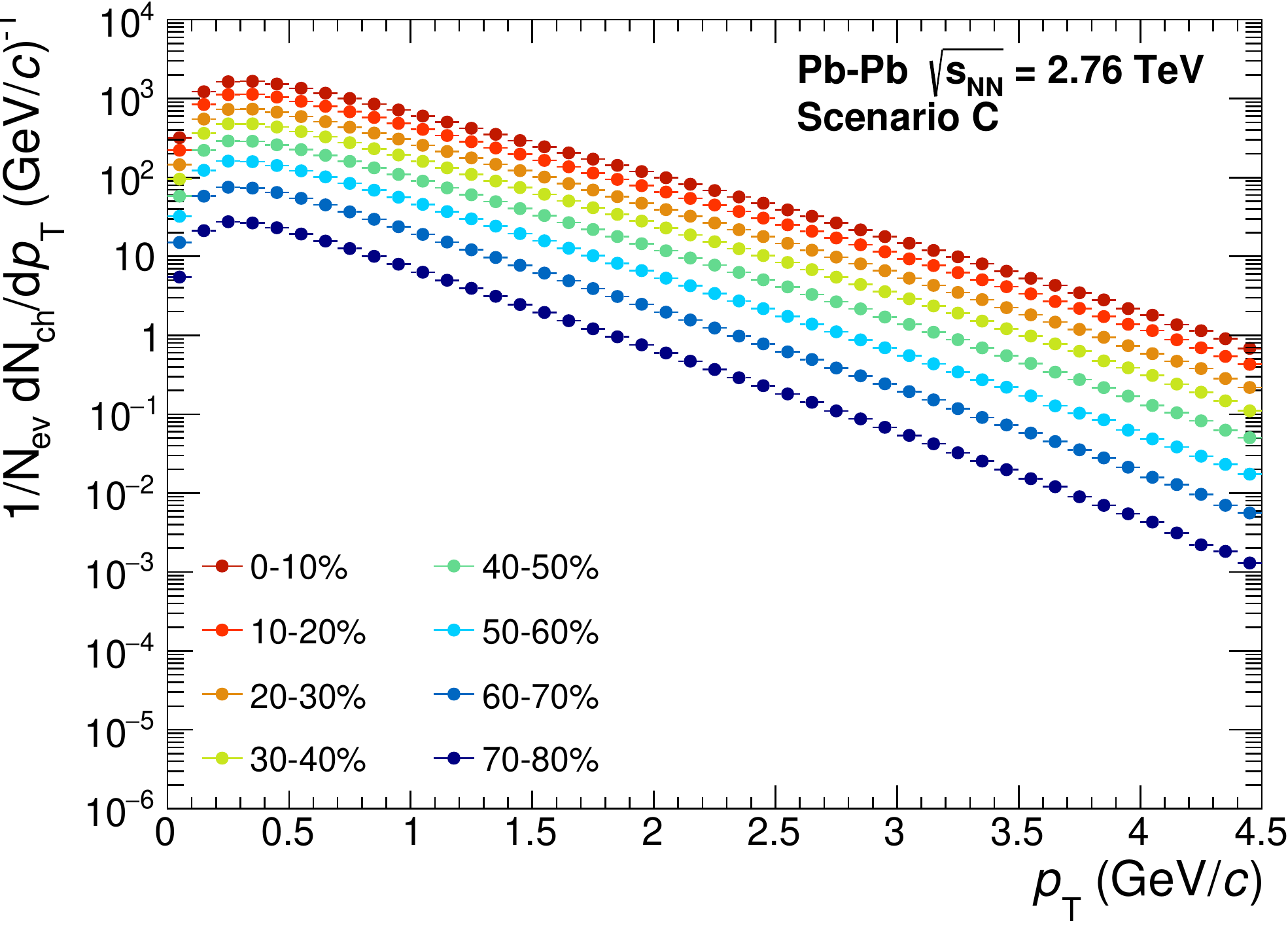}
  \caption{Spectra of momentum of charged particles in the transverse plane for the three scenarios under consideration: A (left), B (middle) and C (right).}
  \label{fig:momentum}
\end{figure}

\begin{figure}[ht]
  \includegraphics[width=.475\linewidth]{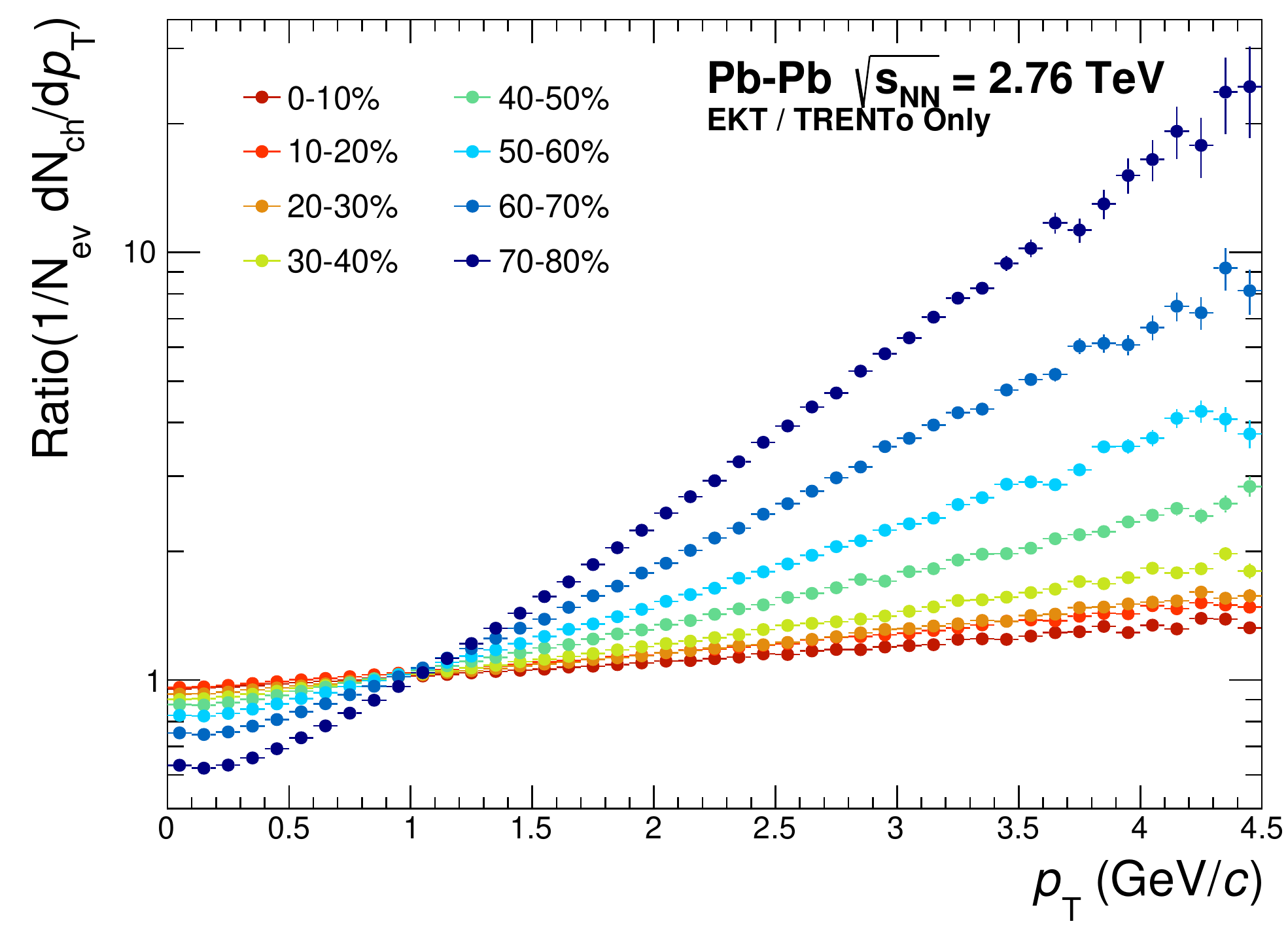}
  \includegraphics[width=.475\linewidth]{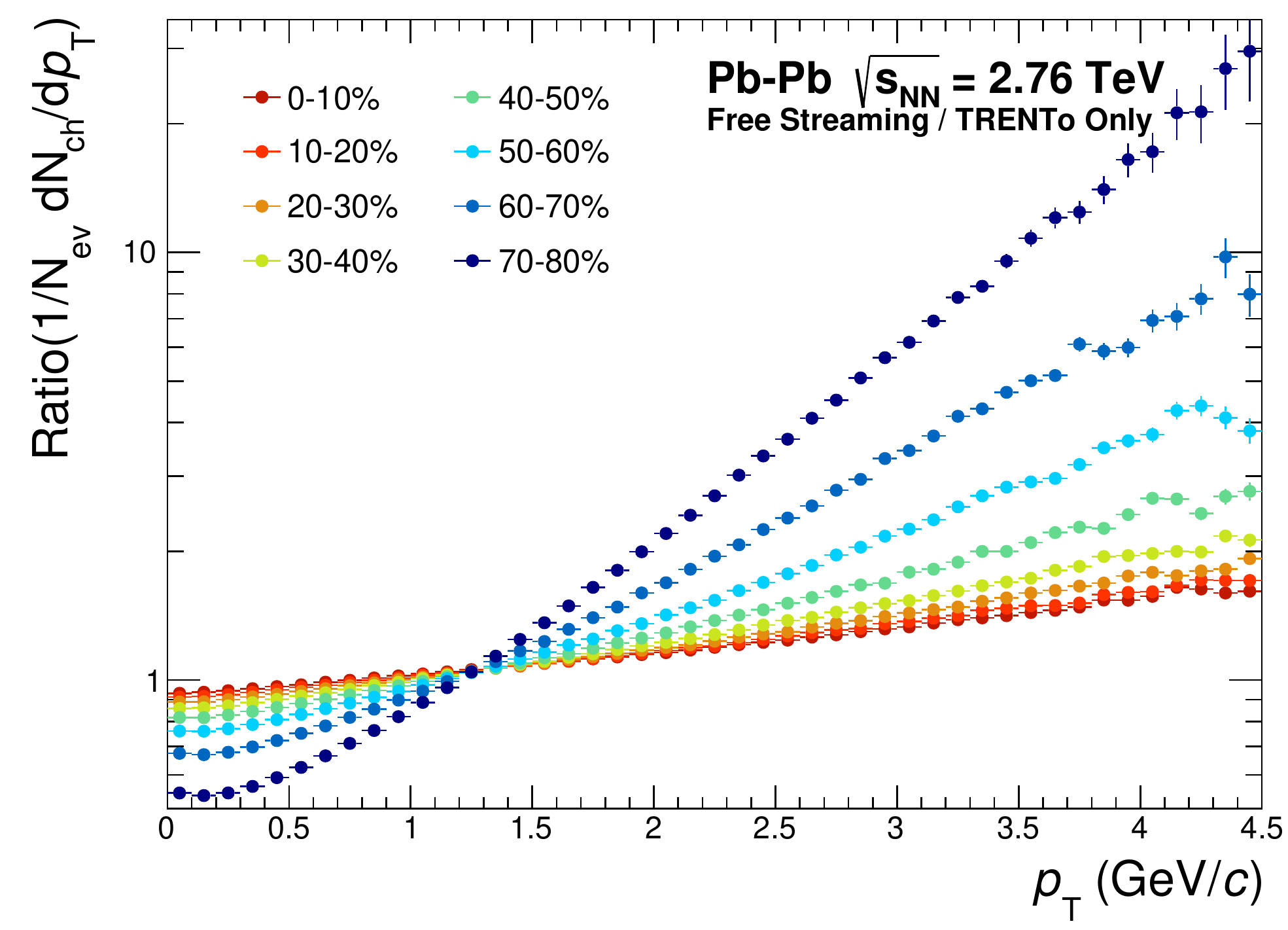}
  \caption{Taking A defined above as the standard scenario, we plot the ratios of the momentum spectra from scenarios B (left) and C (right) with respect to scenario A as a function of the transverse momentum $p_T$.}
  \label{fig:momentum-ratios}
\end{figure}

\begin{figure}
  \includegraphics[width=.55\linewidth]{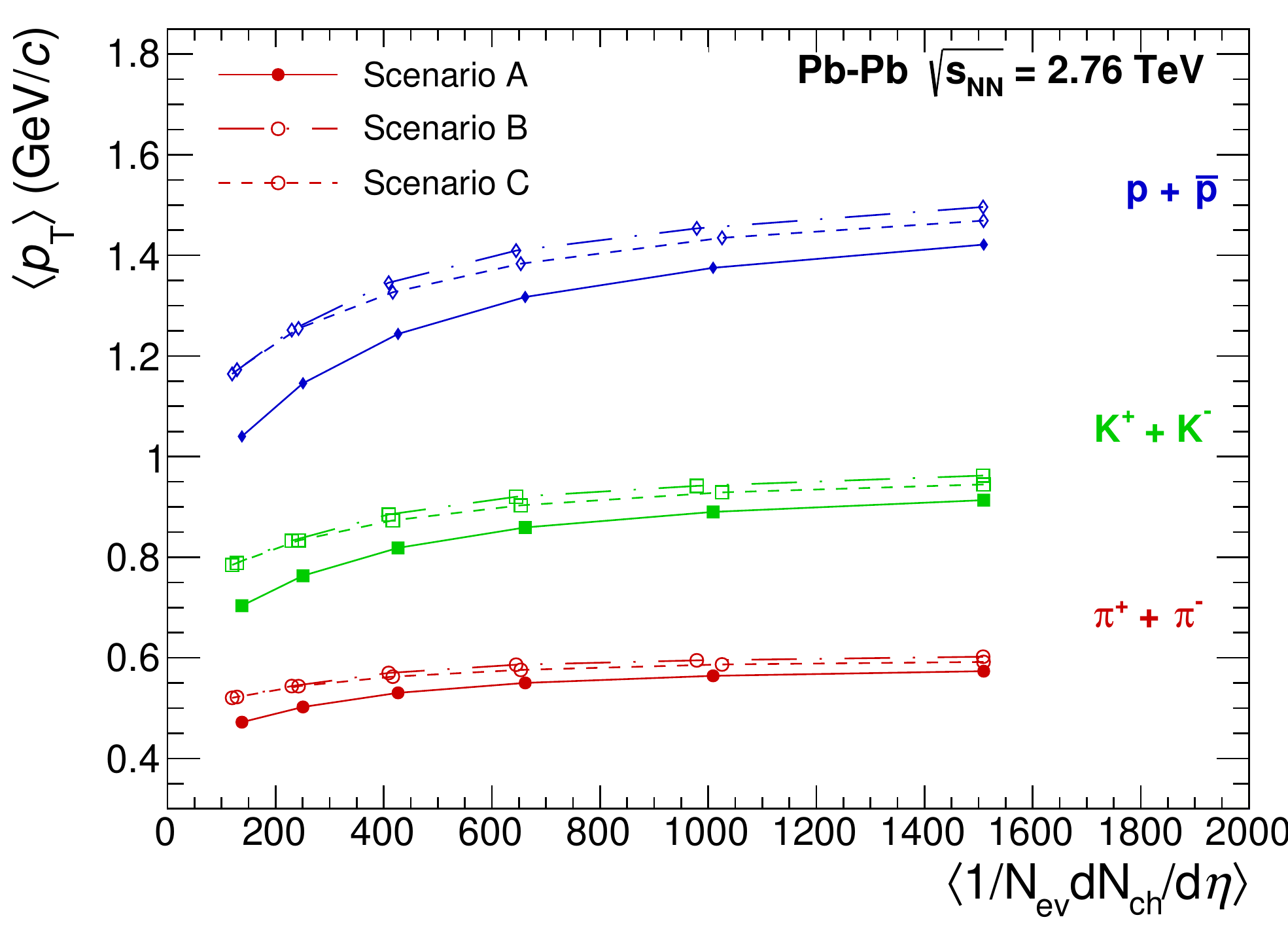}
  \caption{Mean transverse momentum $p_T$ for the three scenarios under investigation and for three sets of particle and anti-particle species: protons (blue), kaons (green), and pions (red), plotted as a function of the multiplicity of charged particles.}
  \label{fig:mean-pT}
\end{figure}

With that in mind, we have calculated the differential $p_T$ spectra of charged particles for the three scenarios under consideration, across several centrality ranges. The results are presented in Fig.~\ref{fig:momentum}, in which it is already possible to notice the effect of the inclusion of pre-equilibrium dynamics on the final momentum spectrum, across all centralities.
In order to make the visualization of this effect clearer, we also show in Fig.~\ref{fig:momentum-ratios} ratios between scenarios B and C and the standard scenario A. 
It is then clear that the addition of K{\o}MP{\o}ST, either in EKT or free streaming mode, causes a significant change in the shape of the transverse momentum spectra, resulting in a smaller spectrum at low $p_T$ that progressively becomes larger at higher values of $p_T$. The relative magnitude of the effect is larger in more peripheral events. Overall, the net result of the effect is an increase in average transverse momentum, which is shown, per centrality class, in Fig.~\ref{fig:mean-pT} for three particle species (and their corresponding anti-particles): pions, kaons, and protons.
This increase in $\langle p_T \rangle$ is ultimately reflected in integrated anisotropic flow observables, explaining the effect observed in Fig.~\ref{fig:anisotropic-integrated}. 

Therefore, both the free streaming limit and the EKT scenario behave similarly, with larger mean transverse momenta compared to the hydrodynamic scenario.  This is despite the fact that the EKT scenario corresponds to a shear viscosity of $\eta/s = 0.16$, which is smaller than even the minimum of the temperature-dependent shear viscosity of the hydrodynamic evolution, while the free streaming limit corresponds to a much larger (i.e., infinite) shear viscosity.  One may wonder whether there is a shared  aspect to the free streaming and EKT scenarios that is responsible for this momentum increase.

\begin{figure}[ht]
	\includegraphics[width=.55\linewidth]{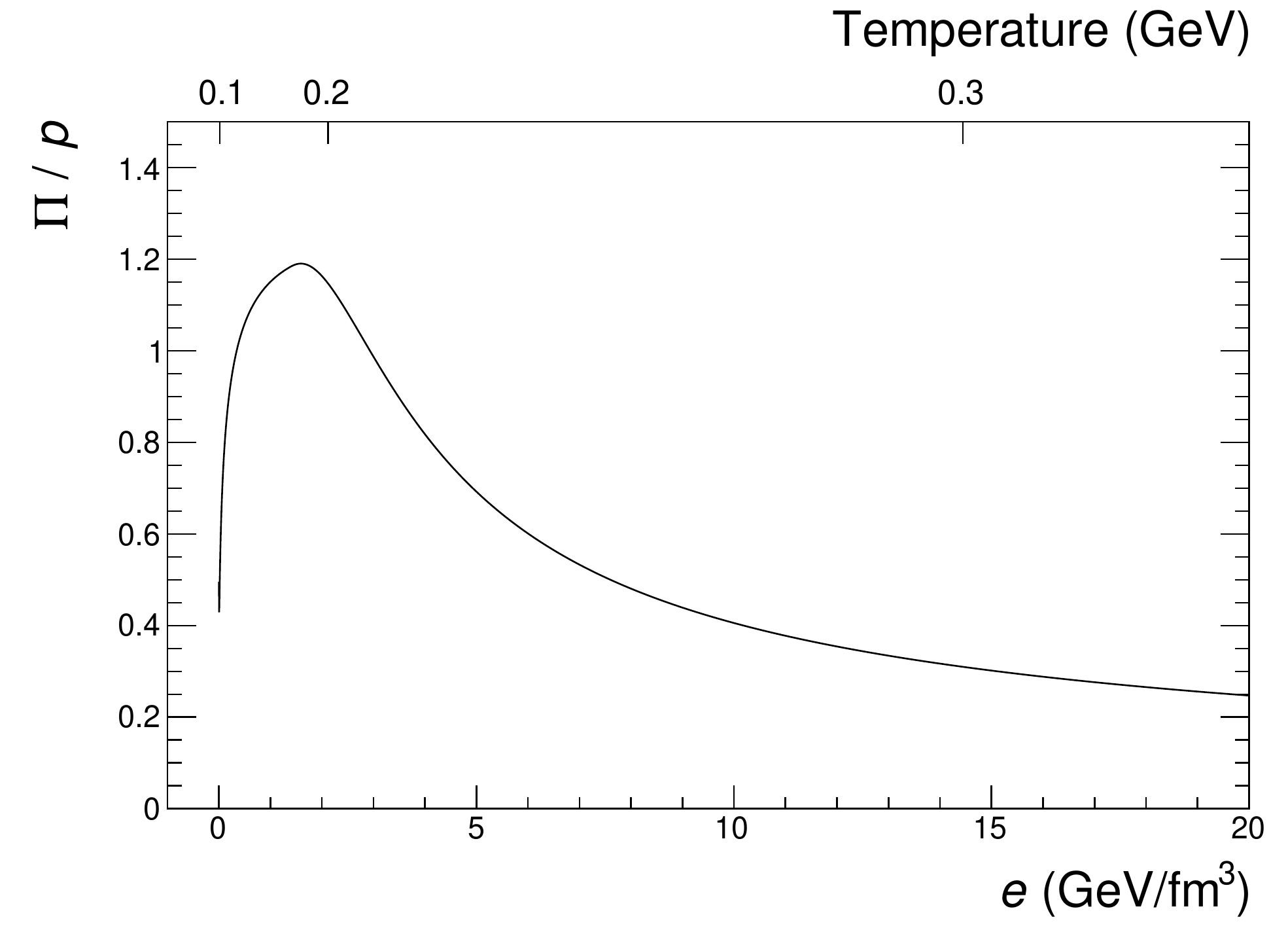}
	\caption{Ratio between the bulk pressure $\Pi$ and QCD thermodynamic pressure $p(e)$ implied by the equation of state s95o-v1.2 combined with the requirement that $T^{\mu}_\mu=0$ at the hydrodynamization time, plotted as a function of the energy density $e$.  The equivalent QCD temperature is also given for reference. See Eq.~\eqref{eqn:TmunuDecompositionConformal}.}
	\label{fig:bulkPressureFromEOS}
\end{figure}

\begin{figure}[ht]
  \includegraphics[width=.475\linewidth]{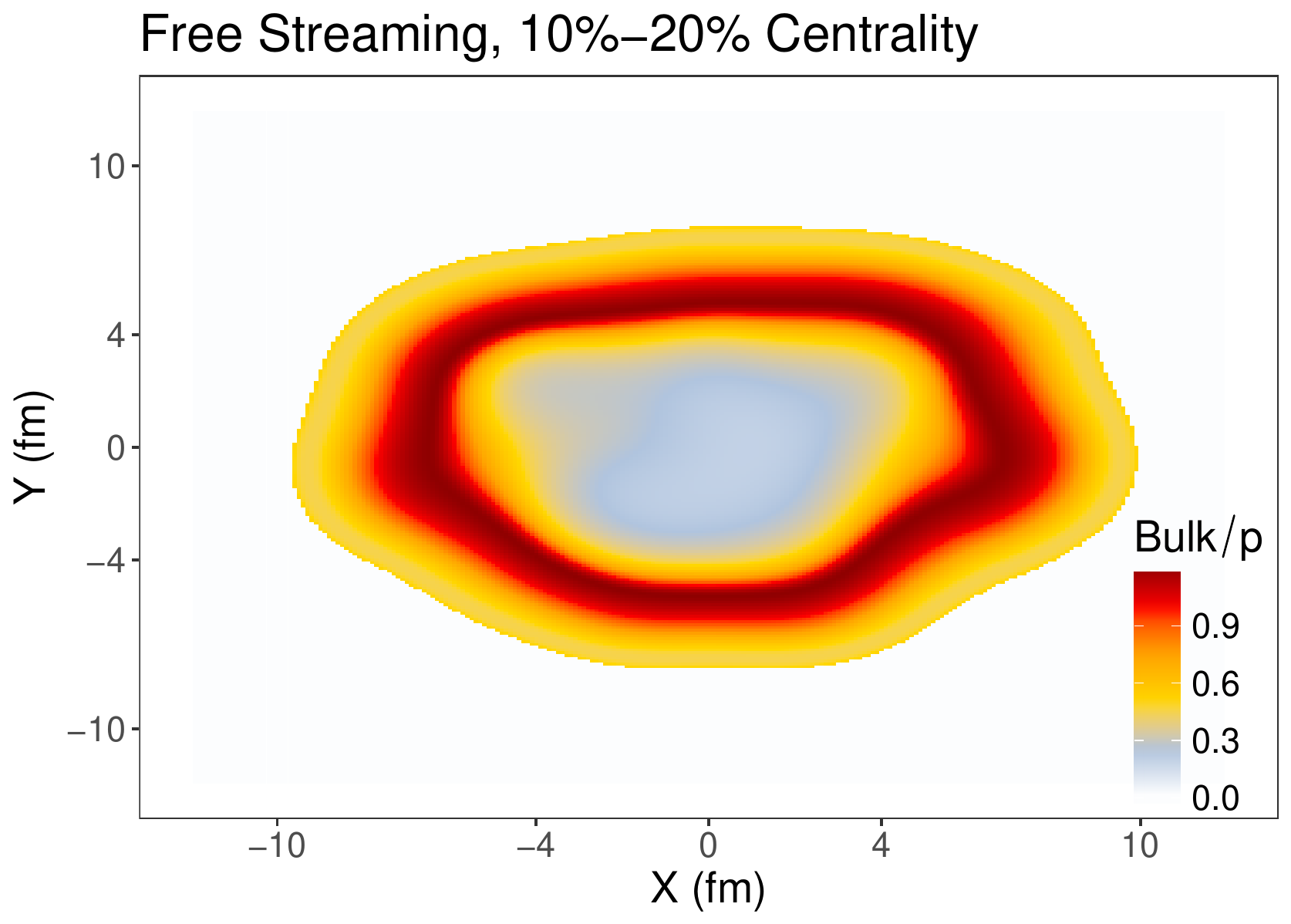}
  \includegraphics[width=.475\linewidth]{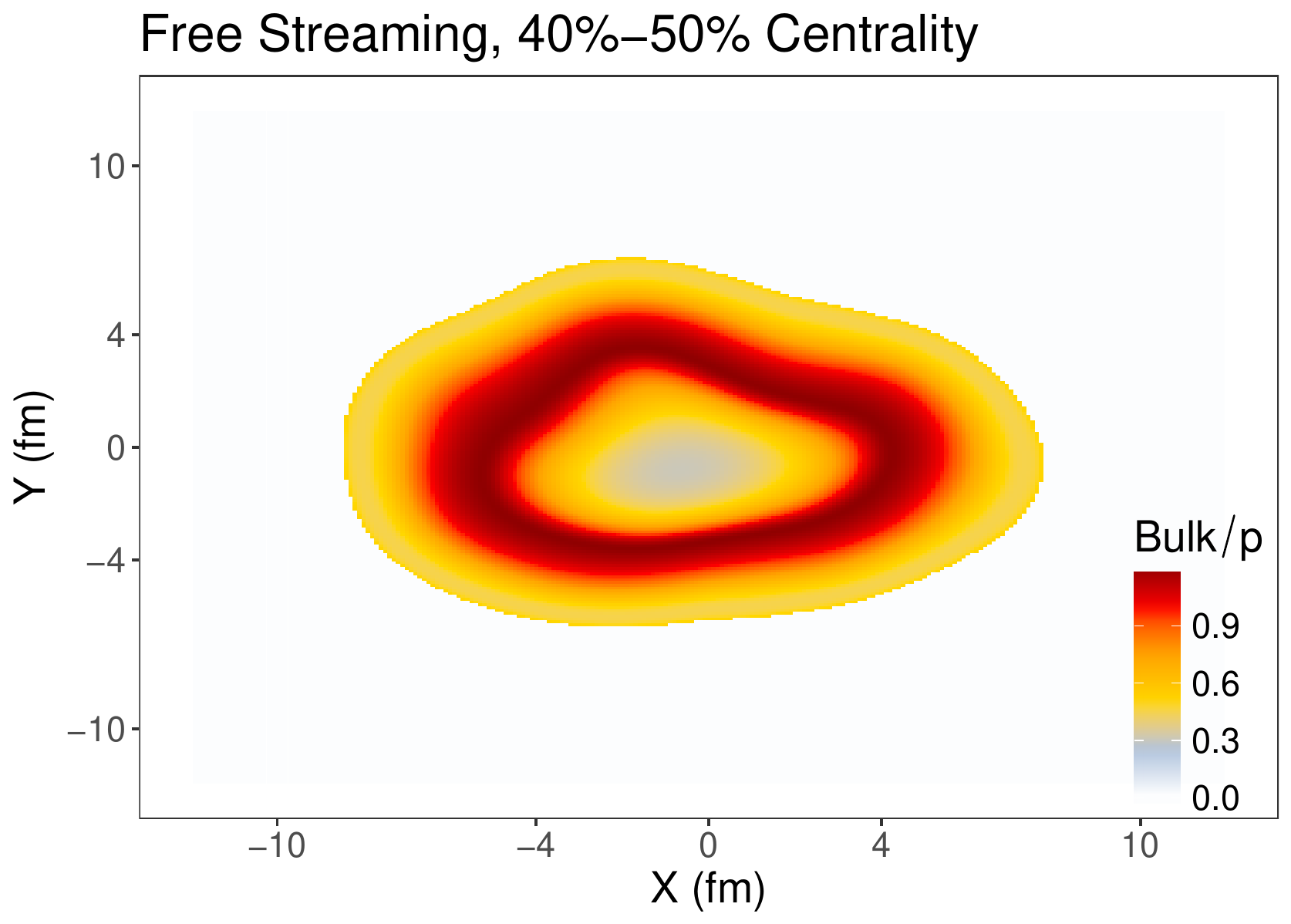}
  \includegraphics[width=.475\linewidth]{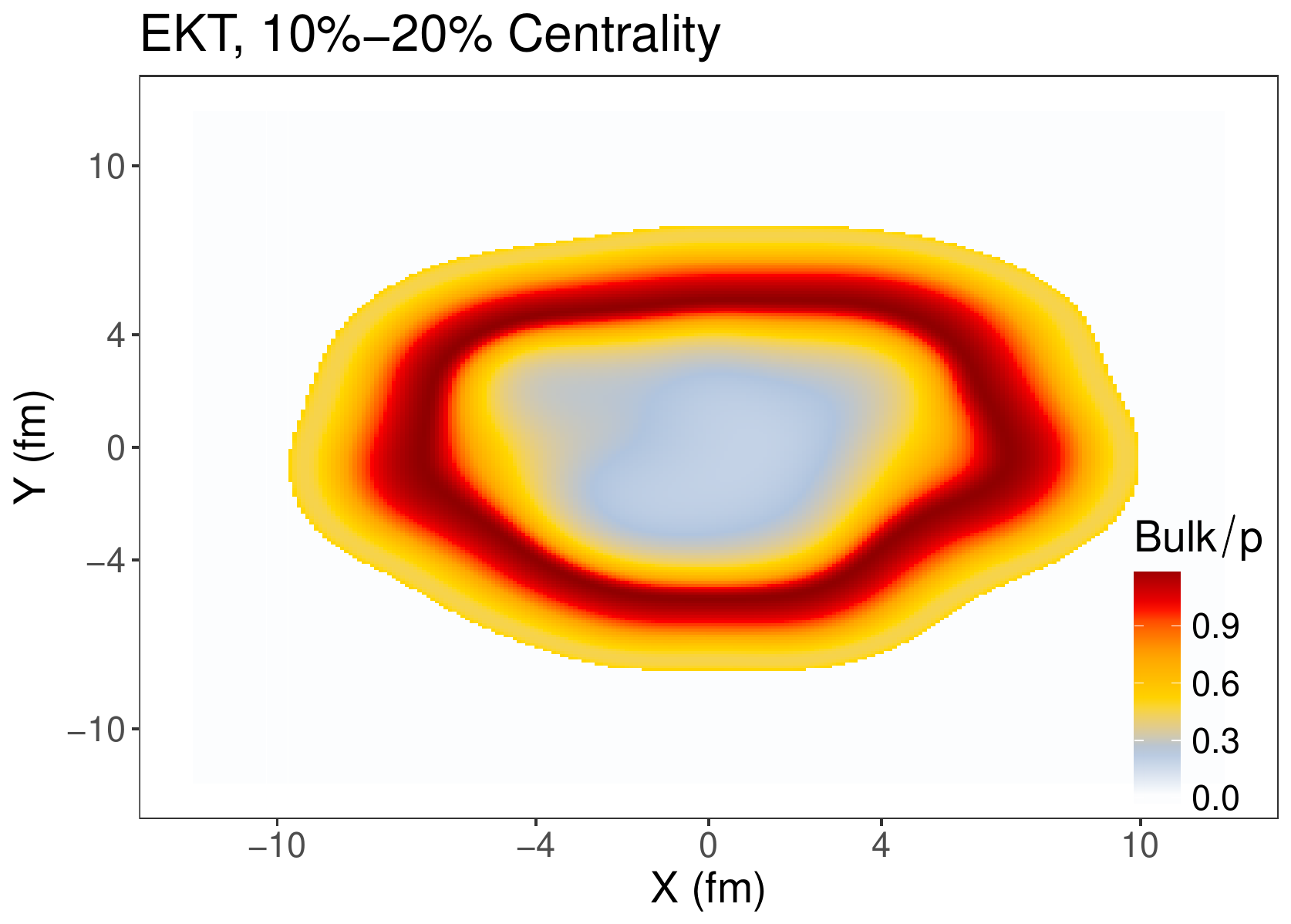}
  \includegraphics[width=.475\linewidth]{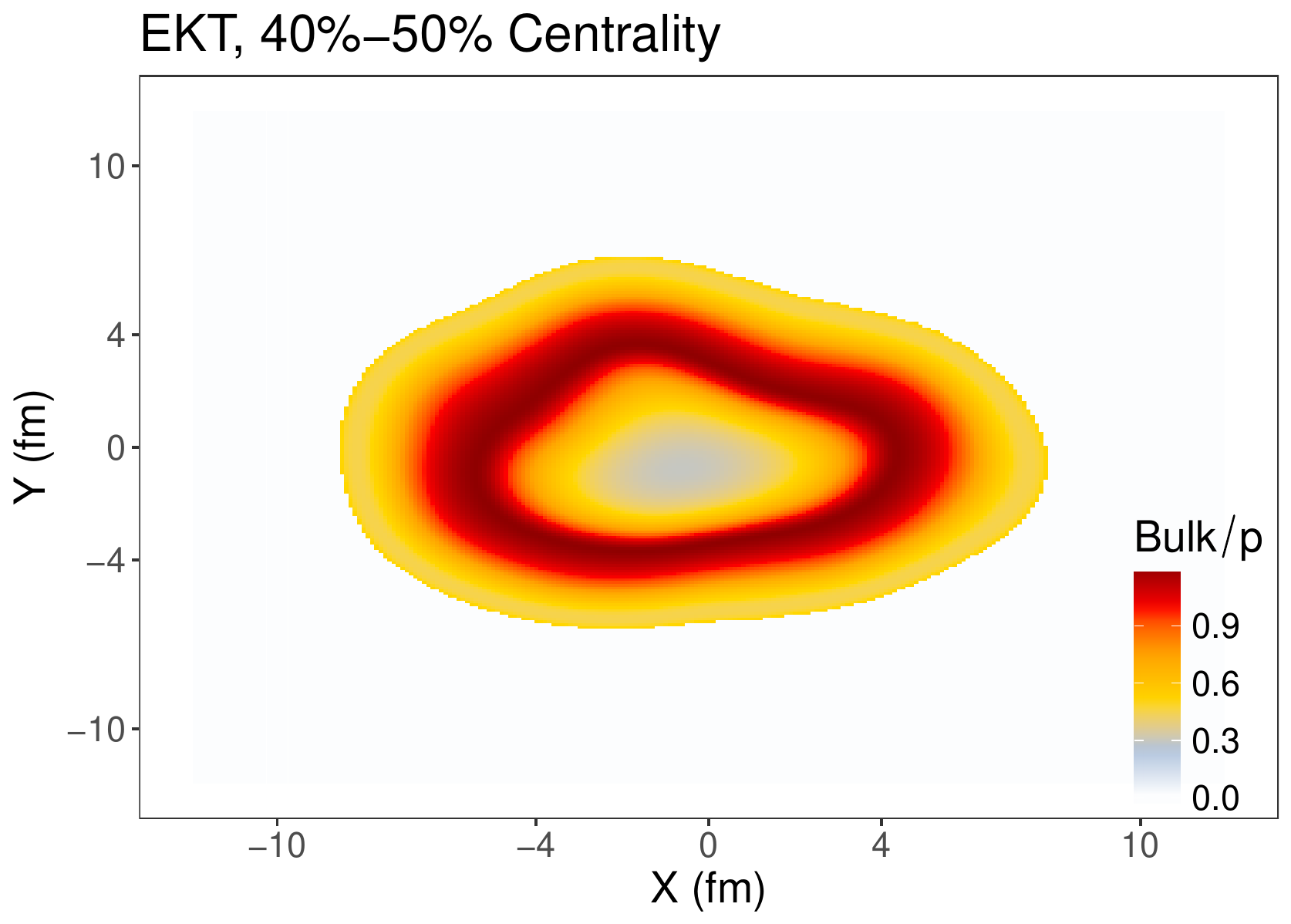}
  \caption{Ratio $\Pi/p$ extracted from sample events in different collision centralities classes and for both the free streaming (top) and EKT (bottom) scenarios at $\tau_\text{hydro}$.}
  \label{fig:bulk-over-p}
\end{figure}

One such possibility is the fact that both scenarios treat the evolution of massless particles.  In fact, this system is conformally invariant and, thus, the trace of the energy-momentum tensor  $T^\mu_\mu$ necessarily vanishes everywhere.  In terms of hydrodynamic variables, this means that the bulk pressure is zero and the total pressure in the kinetic approach is always given by $e/3$. In contrast, it is known that the QCD equation of state is not close to a conformal regime at the temperatures probed in fluid-dynamical simulations of heavy-ion collisions. If one assumes a continuous transition from a kinetic to a hydrodynamical regime, this discontinuity in the thermodynamic pressure must be compensated by introducing an artificial discontinuity in the bulk pressure (which vanished exactly in K{\o}MP{\o}ST). This means that the hydrodynamic evolution is always initialized with a positive bulk viscous pressure $\Pi$. That is, continuity of the energy-momentum tensor combined with a discontinuity of the thermodynamic equation of state demands a corresponding discontinuity in the bulk viscous pressure.Specifically, this leads to the relation
\begin{equation}
\Pi+ p(e) = \frac{e}{3}.
\label{eqn:TmunuDecompositionConformal}
\end{equation}
In Fig.~\ref{fig:bulkPressureFromEOS} we show the ratio of bulk pressure to thermodynamic pressure $\Pi/p(e)$ as a function of energy density.  Note that the conformal invariance of the pre-hydrodynamic model implies, via this matching, that the  initial bulk pressure depends only the local energy density, and does not contain any information about the dynamics of the system. One can see that the bulk pressure can reach values larger than the QCD equilibrium pressure.
We illustrate how this translates to our simulated events in Fig.~\ref{fig:bulk-over-p}, which shows that the bulk pressure at $\tau_{\text{hydro}}$ indeed reaches large positive values, especially near the edge of the system and in the majority of peripheral collision systems.
This positive bulk pressure can indeed increase the radial expansion and, as a result, lead to larger mean transverse momentum. However, we note that this particular aspect of the K{\o}MP{\o}ST model is not realistic. As stated, conformal invariance is not a good approximation for QCD thermodynamics at these temperatures. 

The question then arises how important is this unphysical aspect, and to what extent does it explain the observed change in $p_T$ spectra?   If it is responsible for a significant part of the observed  increase in mean $p_T$, this presents a significant problem, since the change in mean transverse momentum is the main notable effect of pre-hydrodynamic evolution.  

\begin{figure}[ht]
  \includegraphics[width=.475\linewidth]{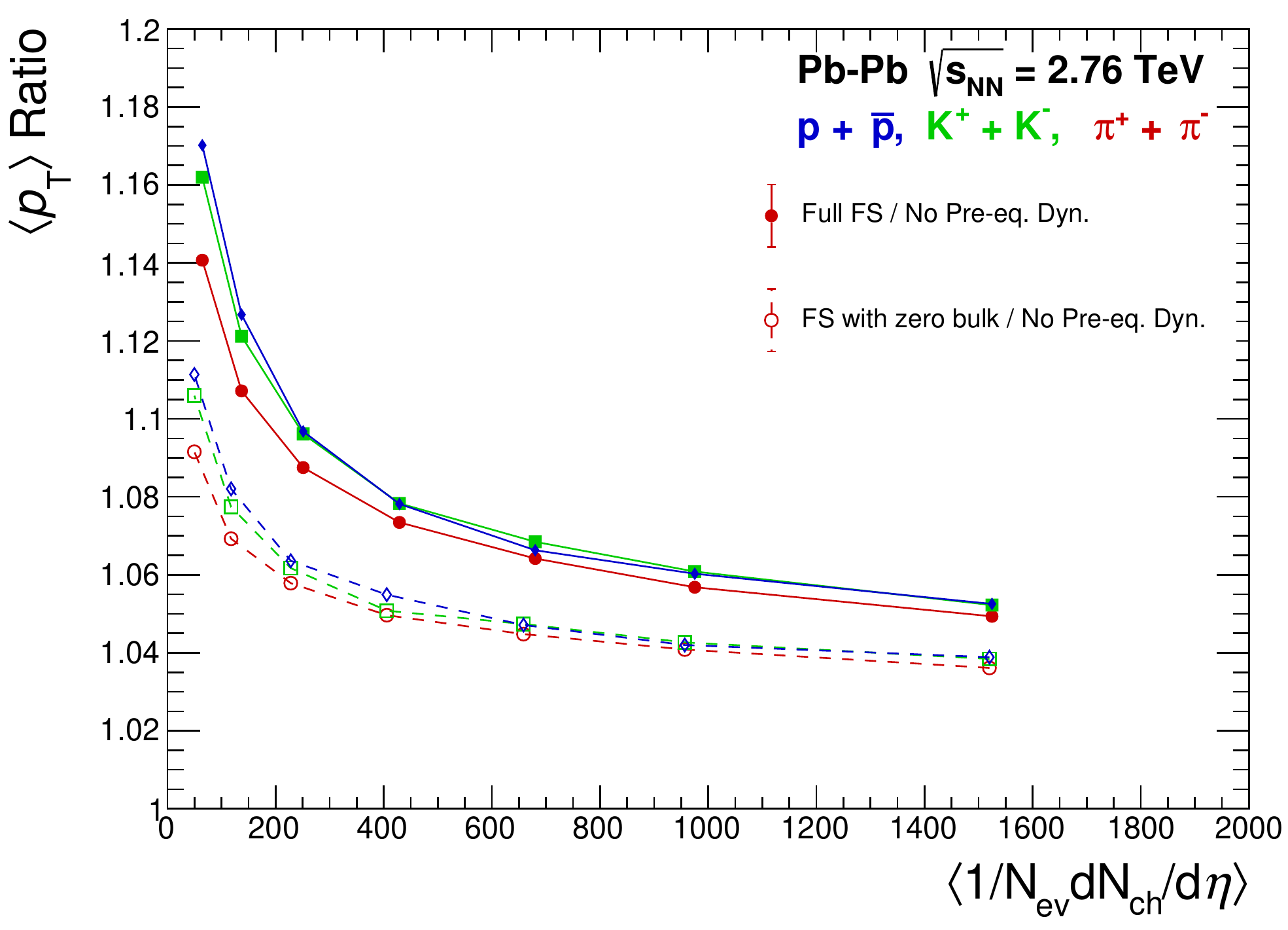}
  \includegraphics[width=.475\linewidth]{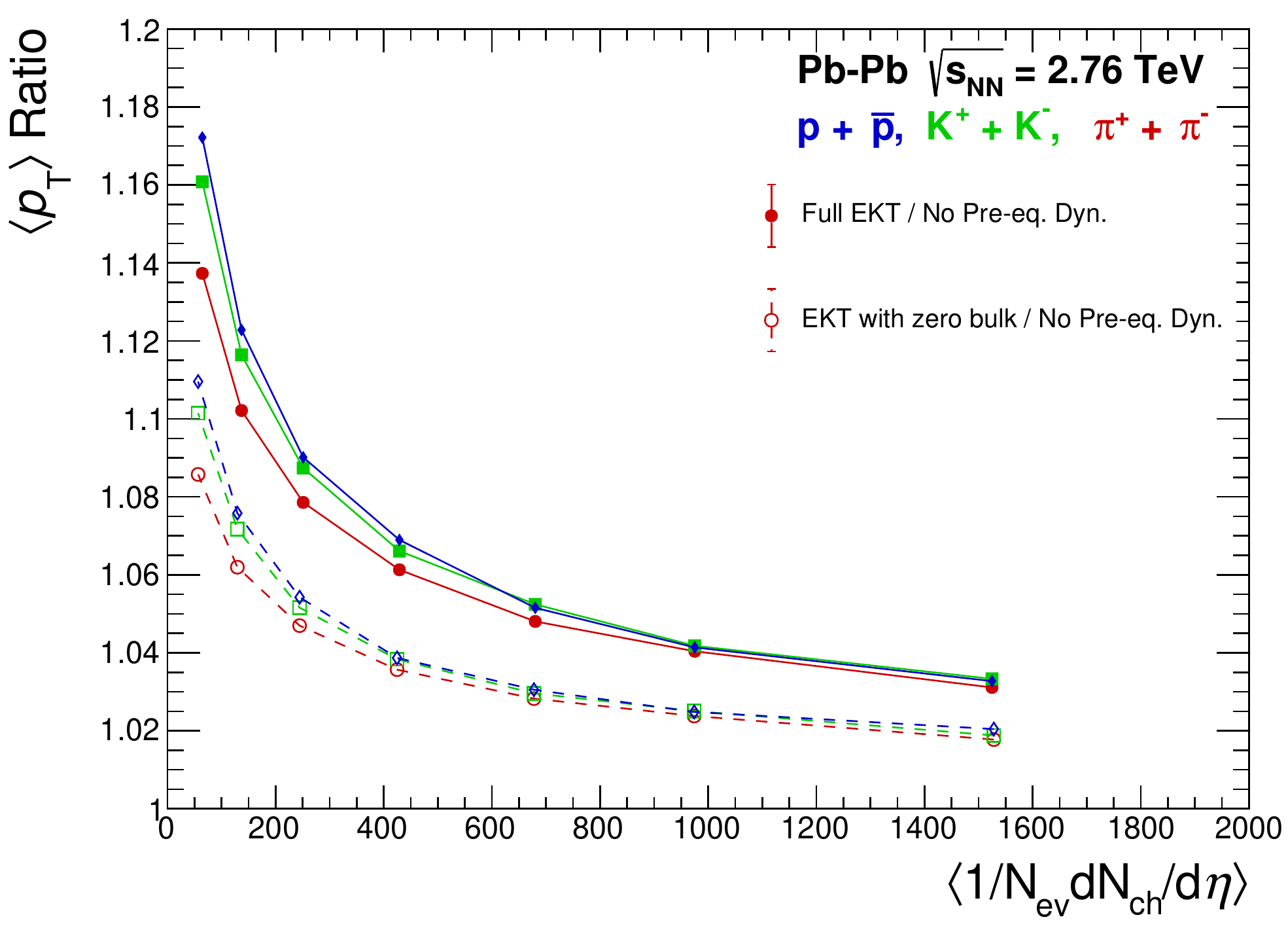}
  \caption{Ratios of mean transverse momentum between scenarios with (painted symbols) and without (blank symbols) initial bulk pressure and the case without pre-equilibrium dynamics, plotted as a function of the charged particle multiplicity, for both Free Streaming (left) and EKT (right).}
  \label{fig:meanpT-zeroBulkRatios}
\end{figure}

To investigate this, we have performed new simulations, for a subset of 300 of the original initial \trento profiles, for scenarios B and C (i.e., with K{\o}MP{\o}ST in free streaming and EKT modes), but ignoring the bulk pressure at the hydrodynamization time (i.e., this quantity is set to zero at the beginning of hydrodynamical evolution). It should be noted that this procedure does not conserve energy and momentum, and we use it only a rough estimate of the effect we want to study. We then compare final results for the mean transverse momentum with the results previously obtained by taking the ratios between both the EKT and free streaming scenarios to the baseline scenarios (without pre-equilibrium dynamics). These ratios are shown in Fig.~\ref{fig:meanpT-zeroBulkRatios}. 
\begin{figure}[ht]
  \includegraphics[width=.475\linewidth]{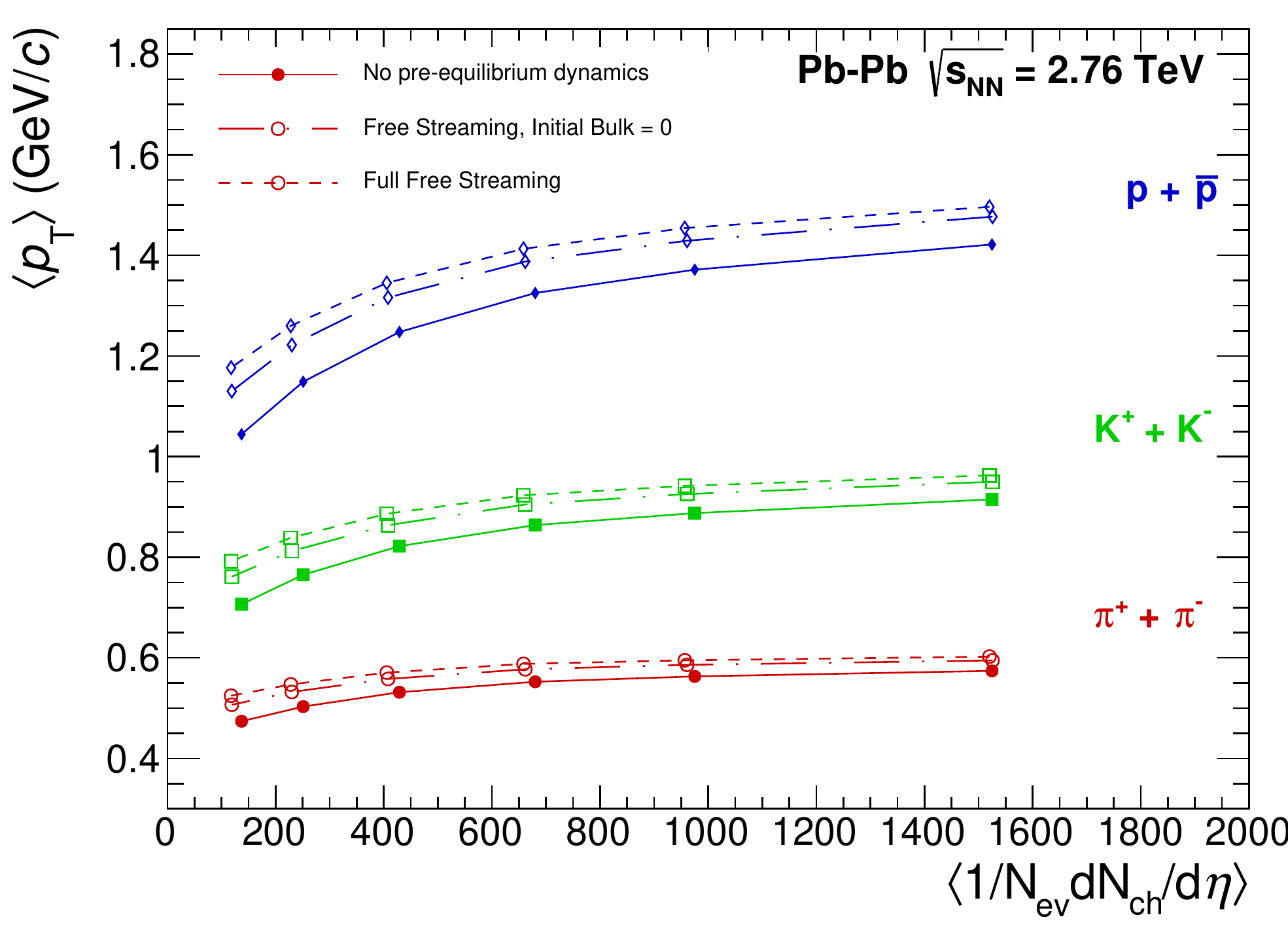}
  \includegraphics[width=.475\linewidth]{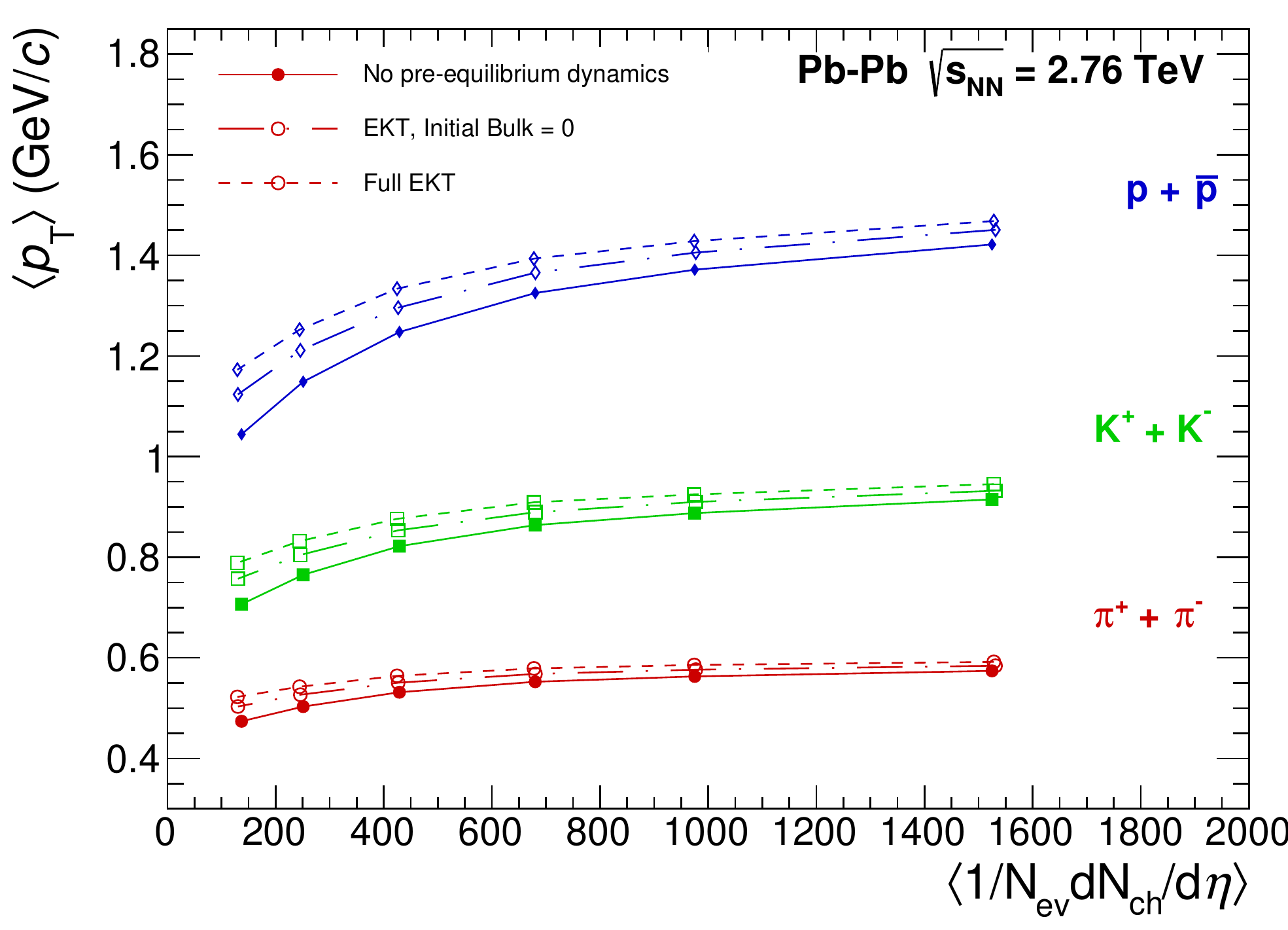}
  \caption{Mean transverse momentum with full pre-equilibrium dynamics and with the bulk pressure set to zero at $\tau_\text{hydro}$, plotted as a function of the charged particle multiplicity, for both free streaming (left) and EKT (right) scenarios. The scenario with no pre-equilibrium dynamics is also shown as a baseline case.}
  \label{fig:meanpT-zeroBulk}
\end{figure}
Indeed, it appears that a potentially large fraction of the increase in $p_T$ may come from the large bulk pressure pressure at $\tau_{\text{hydro}}$. In order to better quantify this effect, we plot the ratios between the scenarios described above and the baseline case without pre-equilibrium dynamics. These are shown in Fig.~\ref{fig:meanpT-zeroBulkRatios}, where is it clear that the fraction of transverse momentum added by the initial bulk pressure pressure is significant and increases for less central events.
The ultimate consequence of this fact is an unphysical enhancement of mean transverse momentum, which is reflected in some of the studied final-state observables as shown in this work.

\section{Conclusions and Outlook}\label{sec:conclusions}
In this work, we have studied how two different scenarios of pre-equilibrium dynamics, namely free streaming and effective kinetic theory implemented via K{\o}MP{\o}ST, affect a collection of final-state observables in relativistic heavy-ion collisions employing a state-of-the-art model of heavy-ion collisions.

We found that the PCA observables, which were devised as a way to study flow fluctuations, are rather insensitive to the details of the pre-equilibrium dynamics and, moreover, to its inclusion in the hybrid model. This strengthens the use of PCA techniques of anisotropic flow as a powerful probe of the hydrodynamic evolution of the system. While the addition of a pre-equilibrium dynamic stage was in general found to be relevant for the calculation of some other observables, we have also found that a potentially large fraction of the observed effects may be an artifact of the underlying assumption of conformal invariance during pre-equilibrium evolution currently implemented in K{\o}MP{\o}ST, which results in a large positive bulk pressure contribution to the pressure at the edges of the resulting initial condition of hydrodynamics at $\tau_\text{hydro}$.  

It will therefore be important in the future to relax the simplifying assumption of conformal invariance in models for the pre-equilibrium stage. We note that this is not a unique property of the K{\o}MP{\o}ST model as it appears in any other model where simplifying approximations are made such that conformal invariance holds. In fact, a similar discussion is valid when using IP-Glasma generated initial conditions \cite{Schenke:2012wb, Schenke:2012fw}. Also, this issue should affect the extraction of transport coefficients from analyses that employ a pre-equilibrium stage described by the evolution of massless partons, such as the Bayesian studies already cited in this work \cite{Bernhard:2018hnz,Bernhard:2019bmu} and a recent study utilising IP-Glasma initial conditions within a hybrid model \cite{gale2020probing}. While the size of the effect depends on the energy density (and therefore on $\tau_{\text{hydro}}$ where the description switches from conformal dynamics to non-conformal hydrodynamics), it is important to keep in mind that this unphysical effect exists when drawing conclusions from comparisons to experimental data using such models.  

Nevertheless, it is important to remark that the general parts of the formalism developed in \cite{Kurkela:2018vqr, Kurkela:2018wud}, which were based on causality and linear response, could in principle be implemented in other microscopic models that are not conformally invariant. In fact, a recent calculation of the Green's function which describes the evolution of energy and momentum perturbations for massless particles in the relaxation time approximation has yielded similar results to those obtained with K{\o}MP{\o}ST \cite{Kamata:2020mka}. Furthermore, in the context of kinetic models one may use a simple gas of particles with temperature dependent masses (in the relaxation time approximation) that can be engineered to describe basic QCD thermodynamic properties, see for instance \cite{Alqahtani:2015qja}. Such a kinetic model would allow for a smooth transition to the hydrodynamic regime where $T^\mu_\mu$ does not vanish at the beginning of the hydrodynamic evolution. However, in this approach one would most certainly lose contact with QCD properties as such models can only be considered, at best, toy models for the non-conformal quark-gluon plasma formed in heavy-ion collisions. 

Keeping in mind the caveats presented above, it is interesting to note that the model we used in this work, using  a set of parameters obtained from a global Bayesian analysis \cite{Bernhard:2018hnz} with the use of a free-streaming pre-equilibrium dynamics stage, except for an overall normalization factor, still yields reasonable final-state observables when compared to experimental data, for a different pre-equilibrium scenario and a different equation of state.

We note that in spite of our rescaling of the \trento profiles, as explained in Section~\ref{sec:setup}, an effect of the pre-equilibrium phase can still be seen as one moves to peripheral events in Figure~\ref{fig:mult}. This is most probably related to the evolution of the longitudinal pressure during this stage, as discussed in \cite{Giacalone:2019ldn}. A detailed discussion of this effect, and its consequences to multiplicity fluctuations, will be deferred to future works. We further intend to explore different models of pre-equilibrium dynamics, aiming at resolving the issues related to the assumption of conformality, in order to provide a clearer picture of hydrodynamization in QCD matter. It is also interesting to explore in detail and quantify how different scenarios of pre-equilibrium dynamics affect the extraction of transport coefficients, and to further confront these scenarios to experimental data. 

\begin{acknowledgments}
We thank J.-F.\ Paquet, A.\ Mazeliuaskas, and Soeren Schlichting for help with the technical numerical aspects within K{\o}MP{\o}ST and for comments on a preliminary version of this manuscript. This research was funded by FAPESP grants number 2016/13803-2 (D.D.C.), 2016/24029-6, 2018/24720-6 (M.L.), 2017/05685-2 (all), 2018/01245-0 (T.N.dS.) and 2018/07833-1 (M.H.). D.D.C., M.L., G.S.D., and J.T. thank CNPq for financial support. G.S.D. acknowledges financial support from Funda\c c\~ao Carlos Chagas Filho de Amparo \`a Pesquisa do Estado do Rio de Janeiro (FAPERJ), grant number E-26/202.747/2018. The authors also acknowledge computing time provided by the Research Computing Support Group at Rice University through agreement with the University of S\~ao Paulo.
\end{acknowledgments}

\appendix
\section*{Appendix: Centrality Dependence of PCA Results}
For the sake of clarity, in this section of present results for some of the observables discussed in the main text for other centrality classes. We start with the PCA results, for the $n=0,2,3$ cases, for centrality classes going from $0-10$\% up to $50-60$\%.
\begin{figure}[ht]
  \includegraphics[width=.32\linewidth]{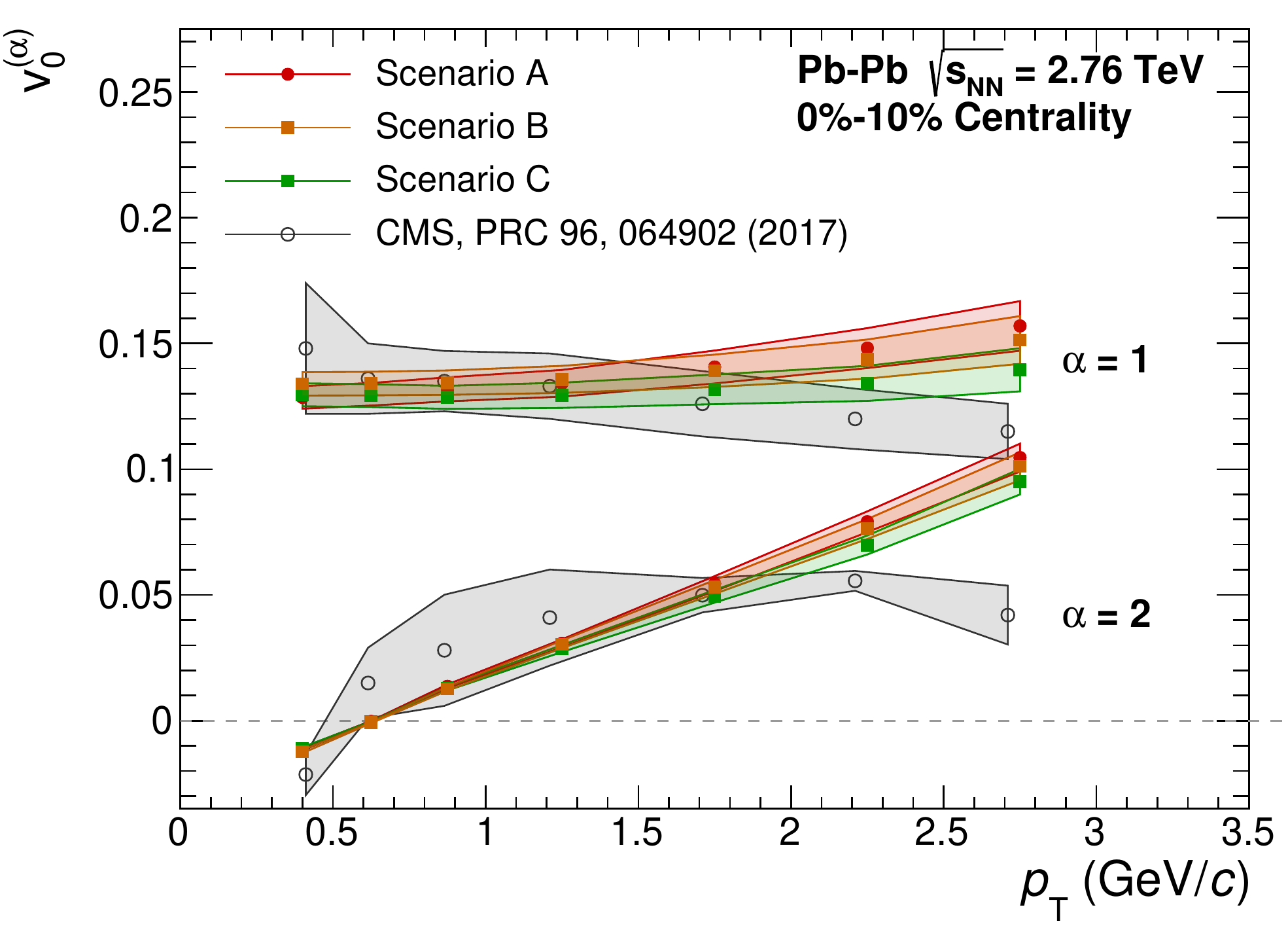}
  \includegraphics[width=.32\linewidth]{figures/pcaBhalerao_n0_10to20.pdf}
  \includegraphics[width=.32\linewidth]{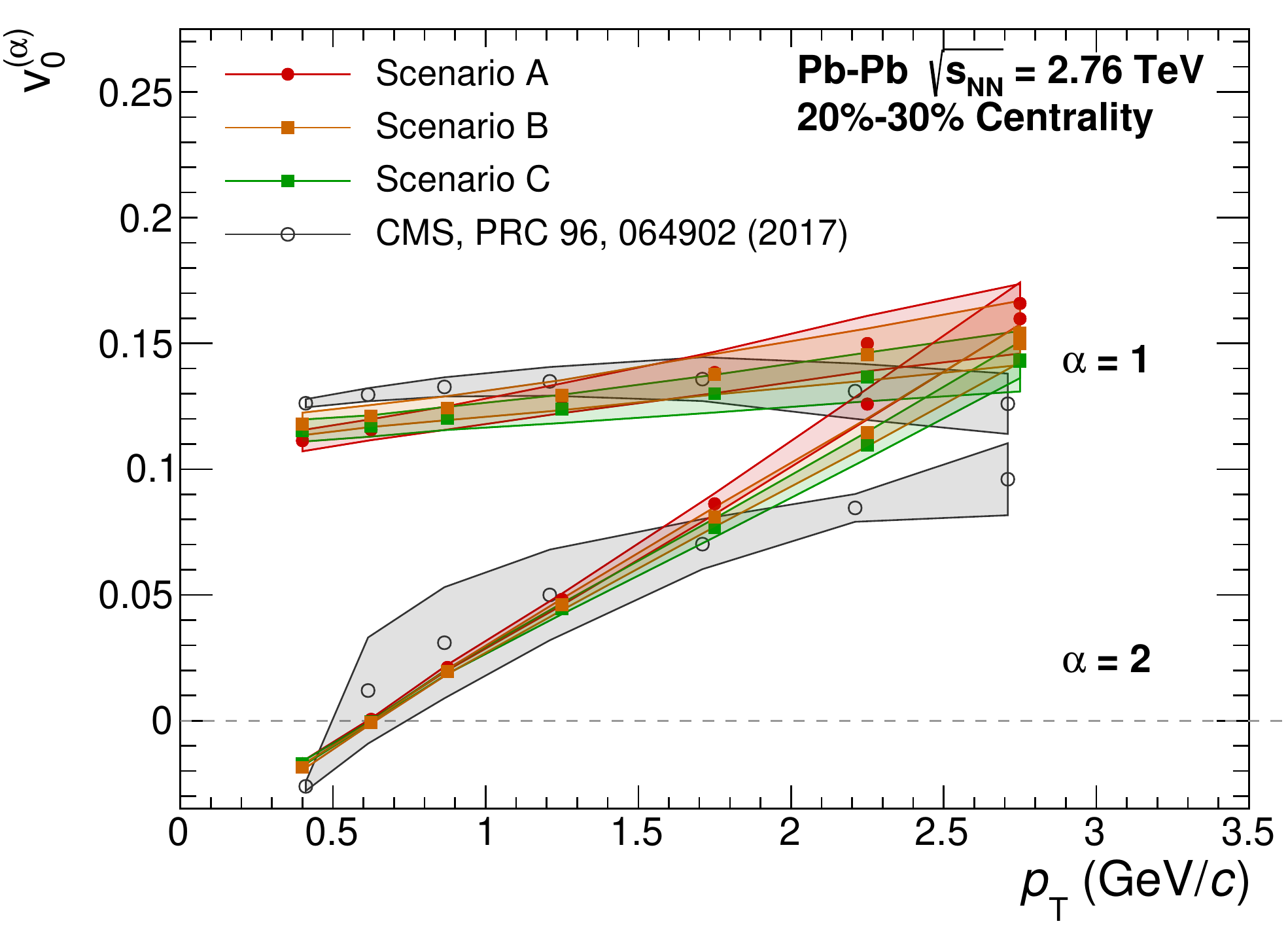}
  \includegraphics[width=.32\linewidth]{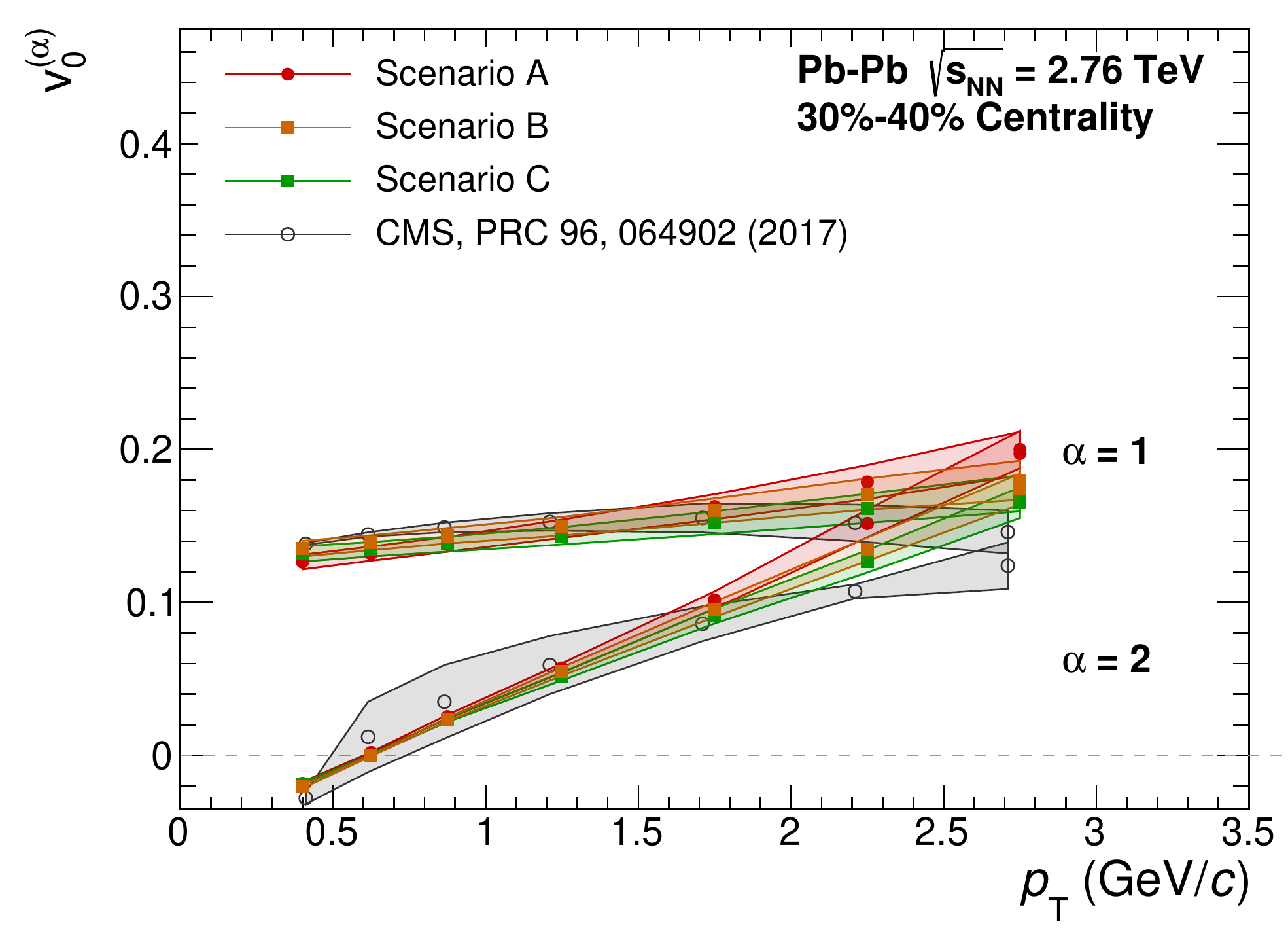}
  \includegraphics[width=.32\linewidth]{figures/pcaBhalerao_n0_40to50.pdf}
  \includegraphics[width=.32\linewidth]{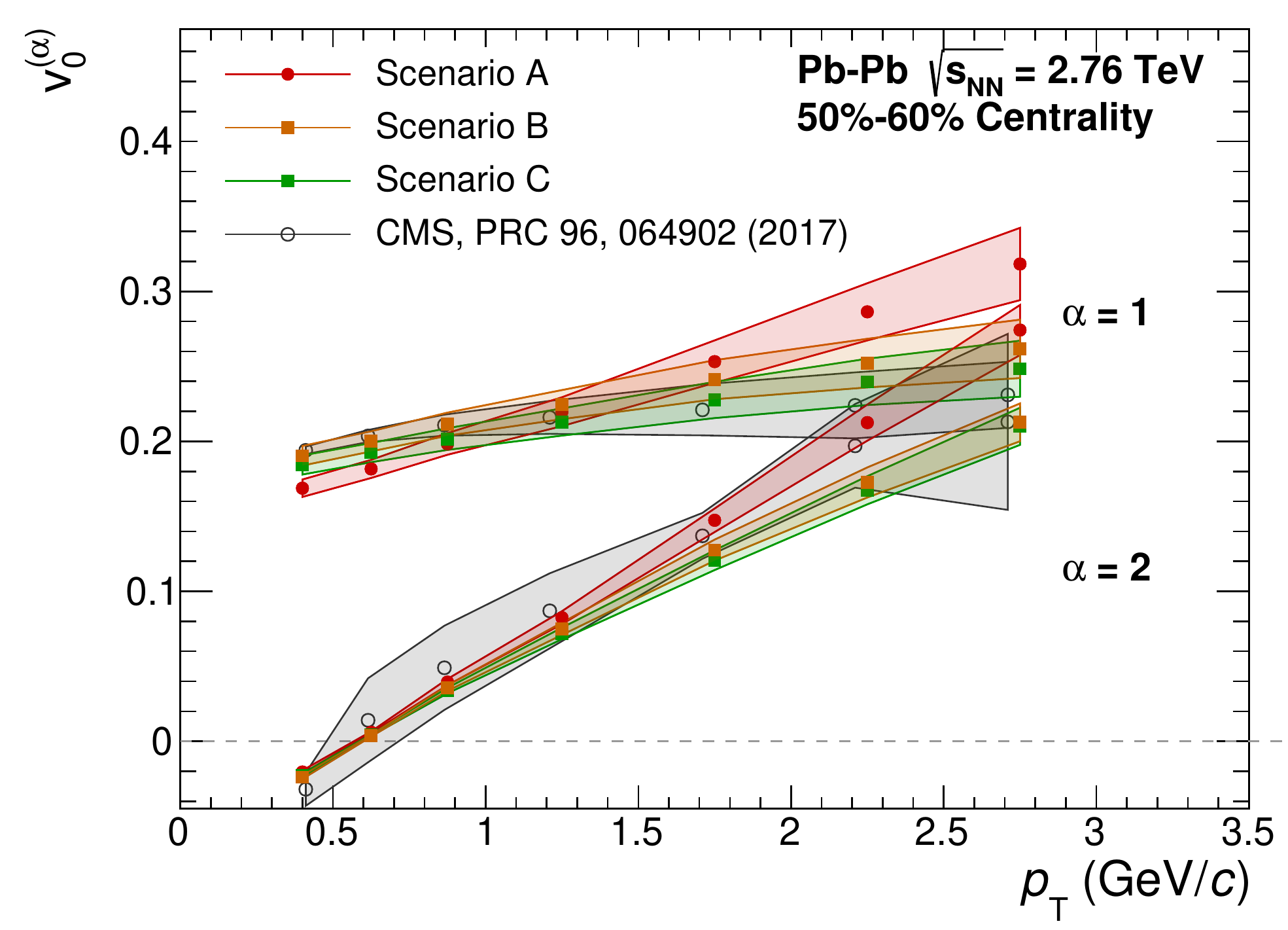}
  \caption{First two principal components of the two-particle correlation matrix according to the prescription by Bhalerao et al. \cite{Bhalerao:2014mua}, plotted as a function of the transverse momentum $p_T$, for the harmonic $n=0$ and several centrality classes.}
  \label{fig:pca0app}
\end{figure}
\begin{figure}[ht]
  \includegraphics[width=.32\linewidth]{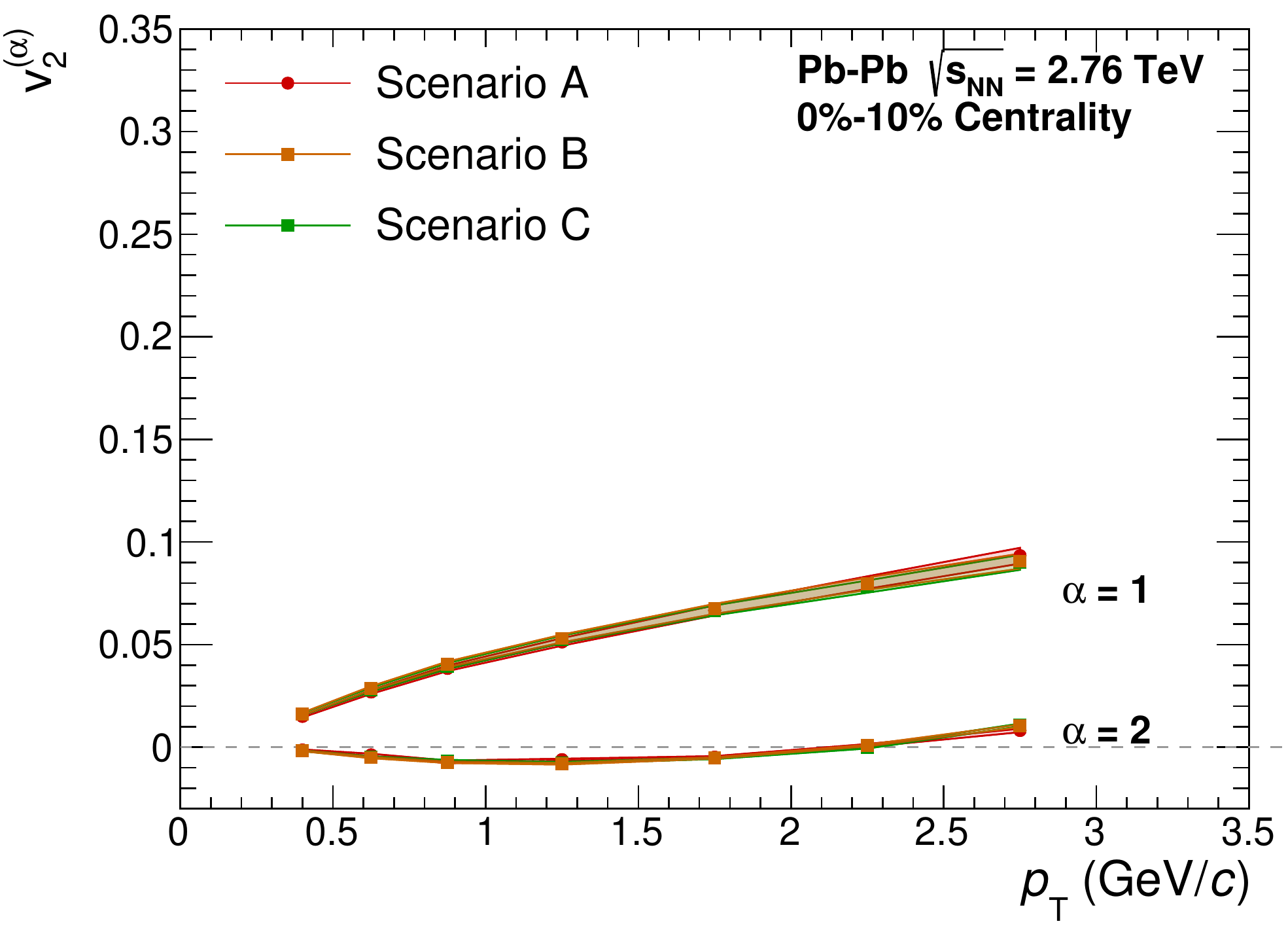}
  \includegraphics[width=.32\linewidth]{figures/pcaExtrem_n2_10to20.pdf}
  \includegraphics[width=.32\linewidth]{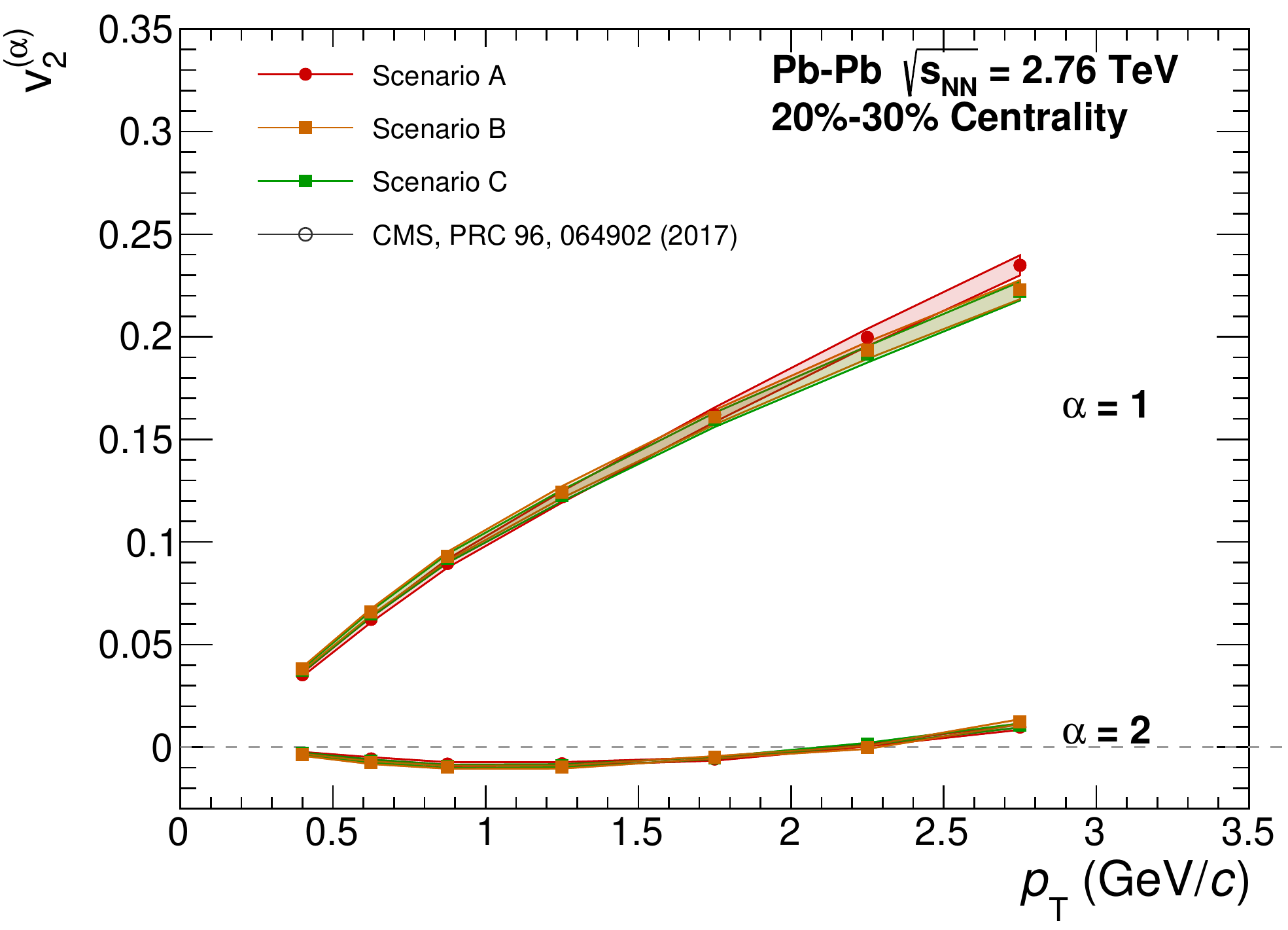}
  \includegraphics[width=.32\linewidth]{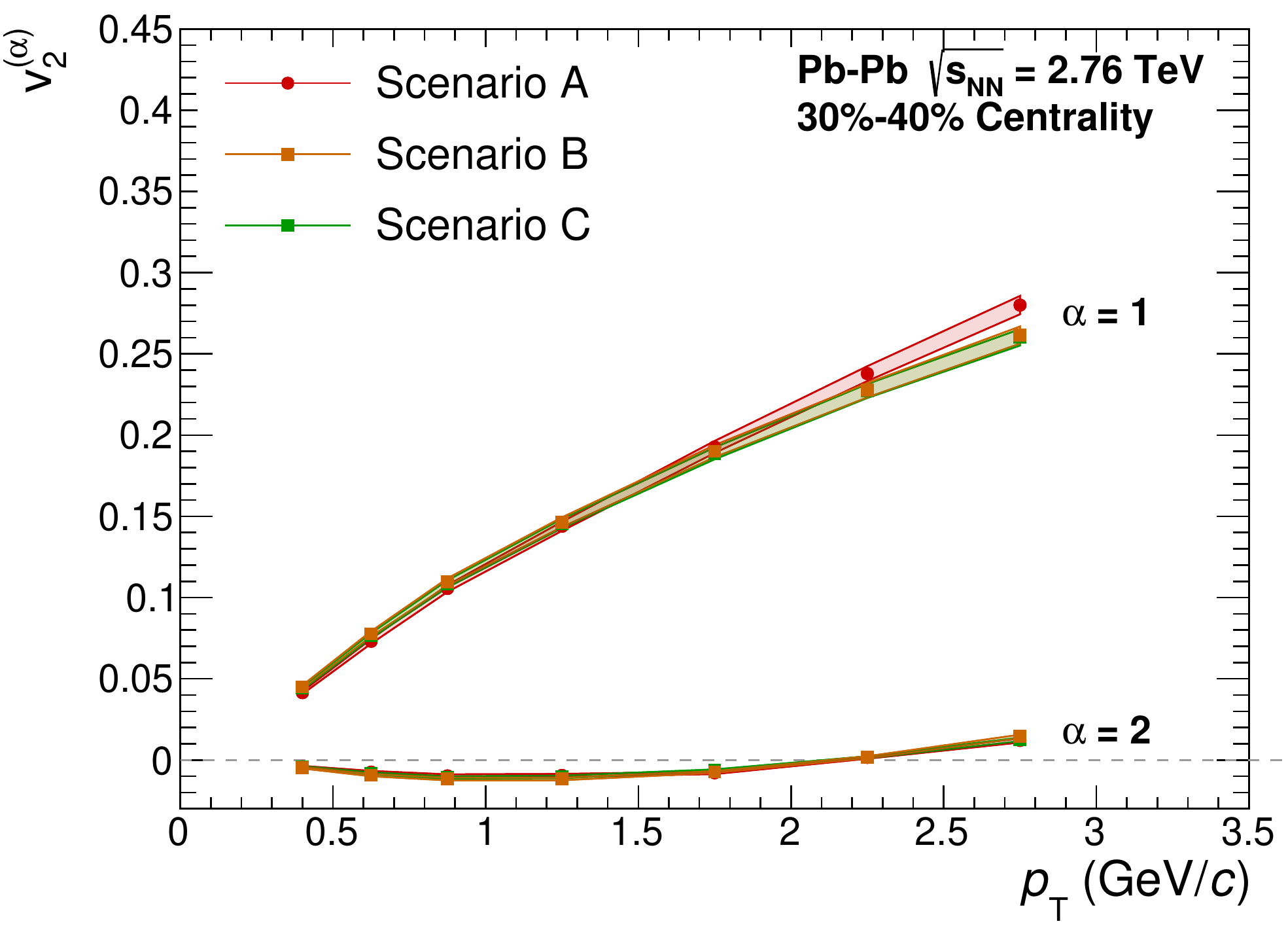}
  \includegraphics[width=.32\linewidth]{figures/pcaExtrem_n2_40to50.pdf}
  \includegraphics[width=.32\linewidth]{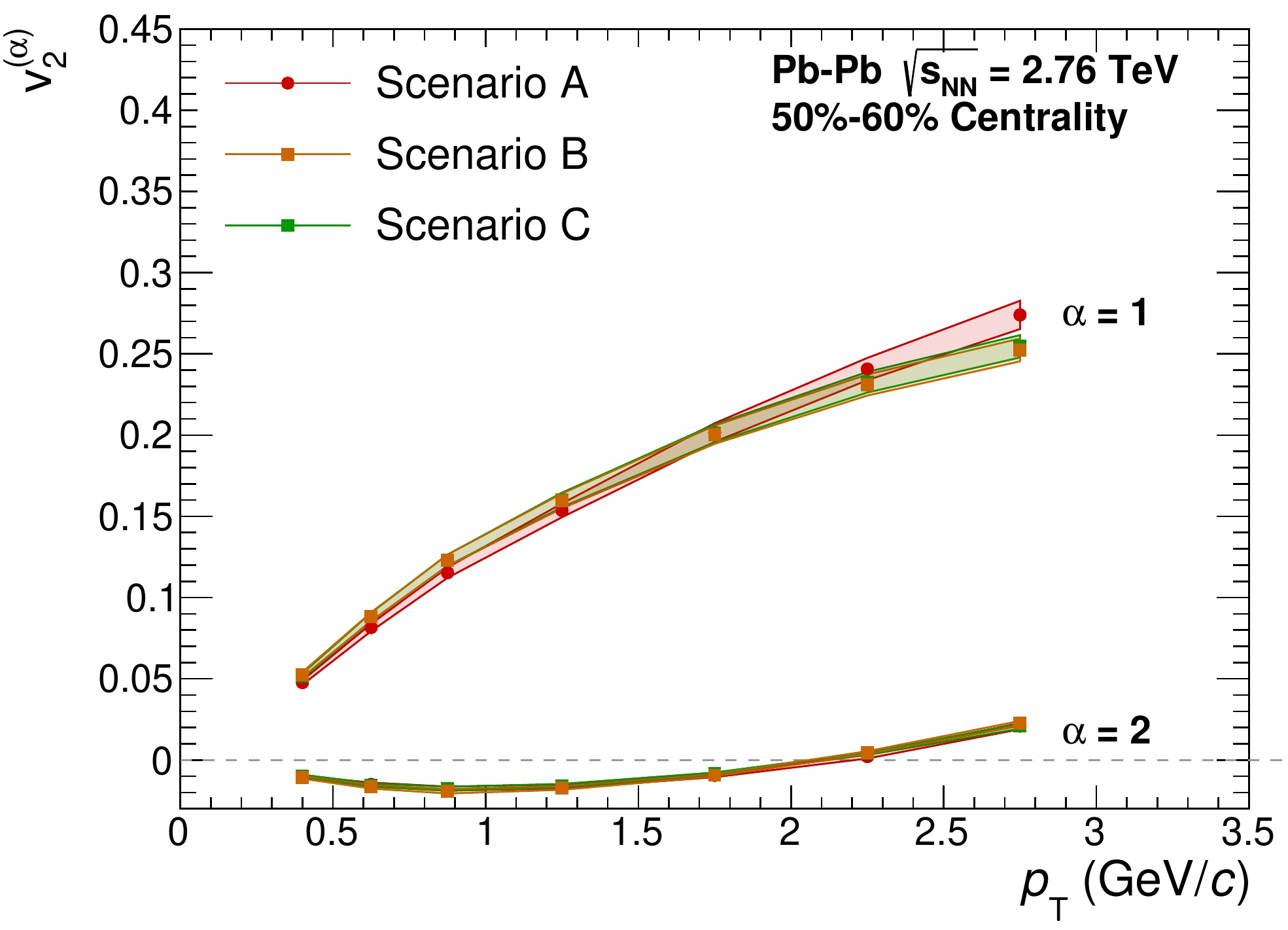}
  \caption{First two principal components of the two-particle correlation matrix according to the ExTrEMe prescription \cite{Hippert:2019swu}, plotted as a function of the transverse momentum $p_T$, for the harmonic $n=2$ and several centrality classes. }
  \label{fig:pca2app}
\end{figure}
\begin{figure}[ht]
  \includegraphics[width=.32\linewidth]{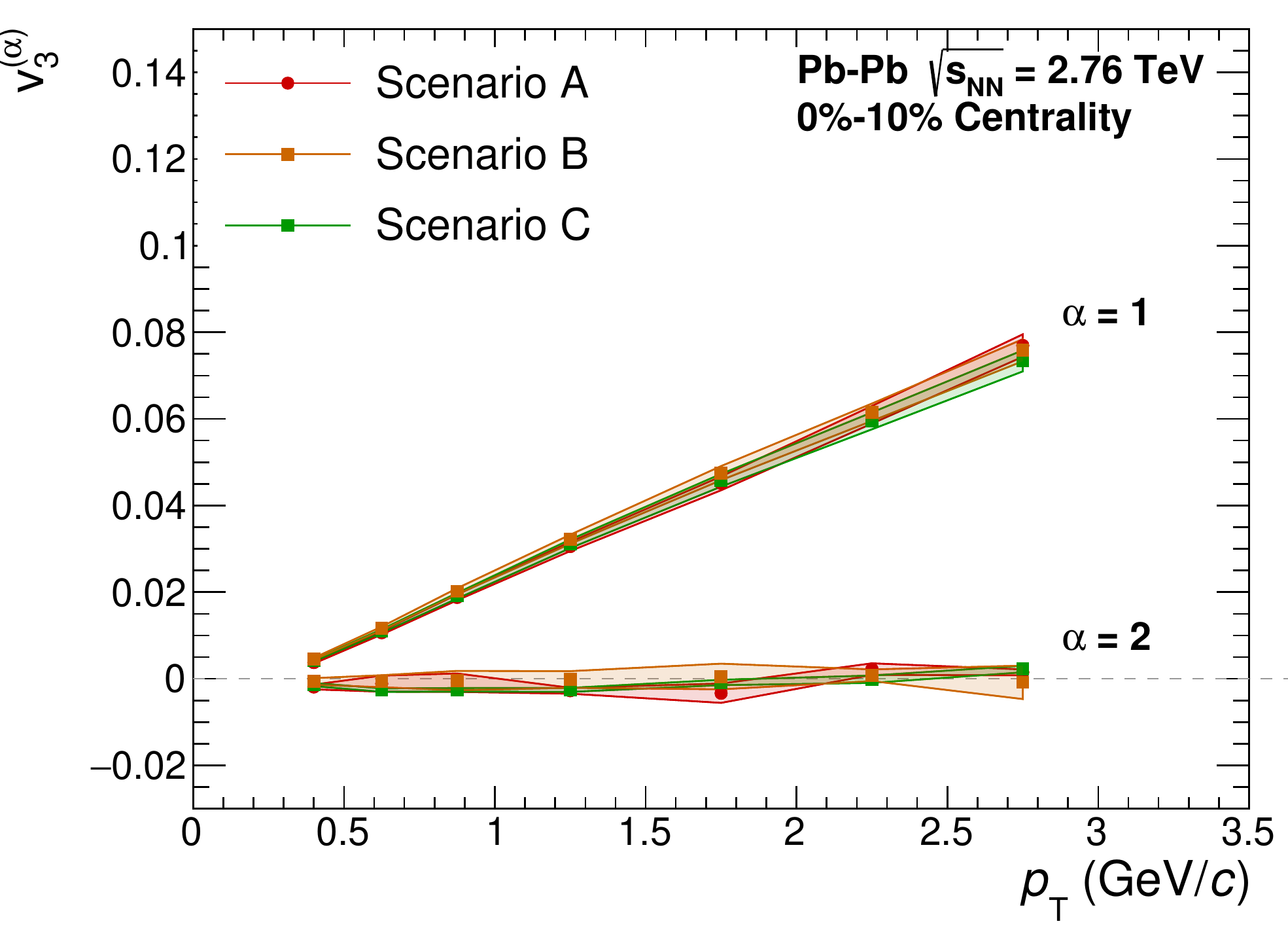}
  \includegraphics[width=.32\linewidth]{figures/pcaExtrem_n3_10to20.pdf}
  \includegraphics[width=.32\linewidth]{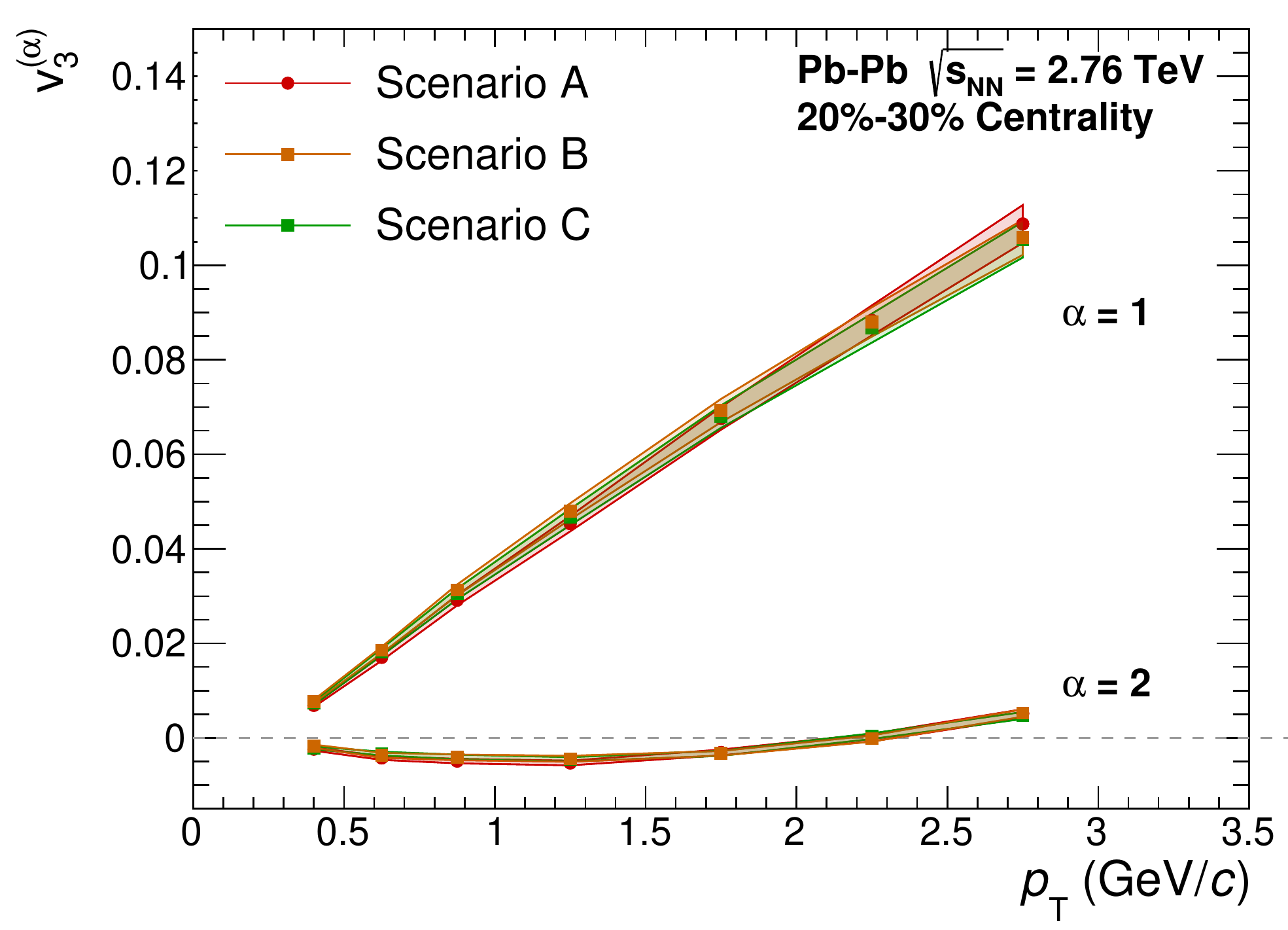}
  \includegraphics[width=.32\linewidth]{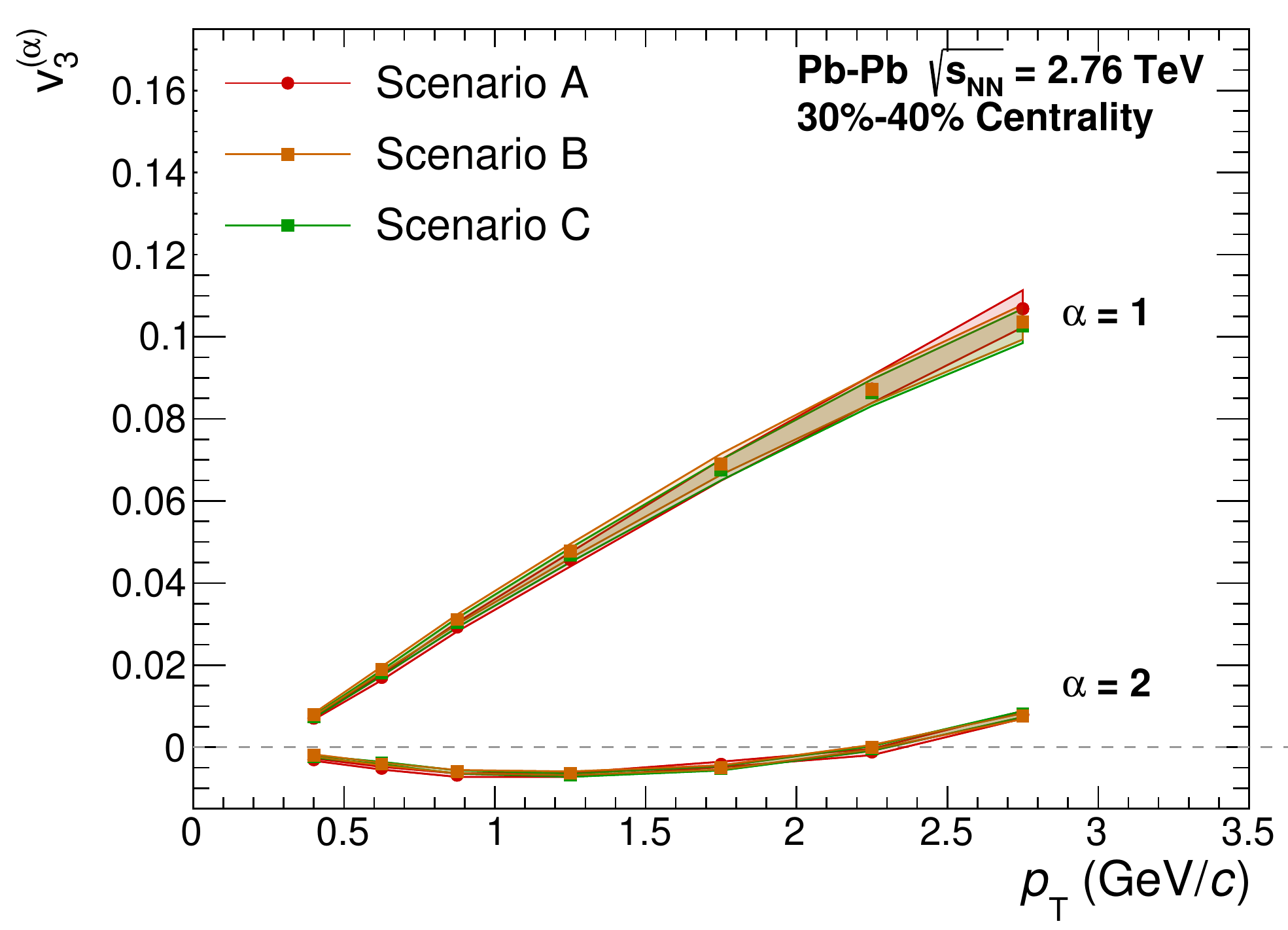}
  \includegraphics[width=.32\linewidth]{figures/pcaExtrem_n3_40to50.pdf}
  \includegraphics[width=.32\linewidth]{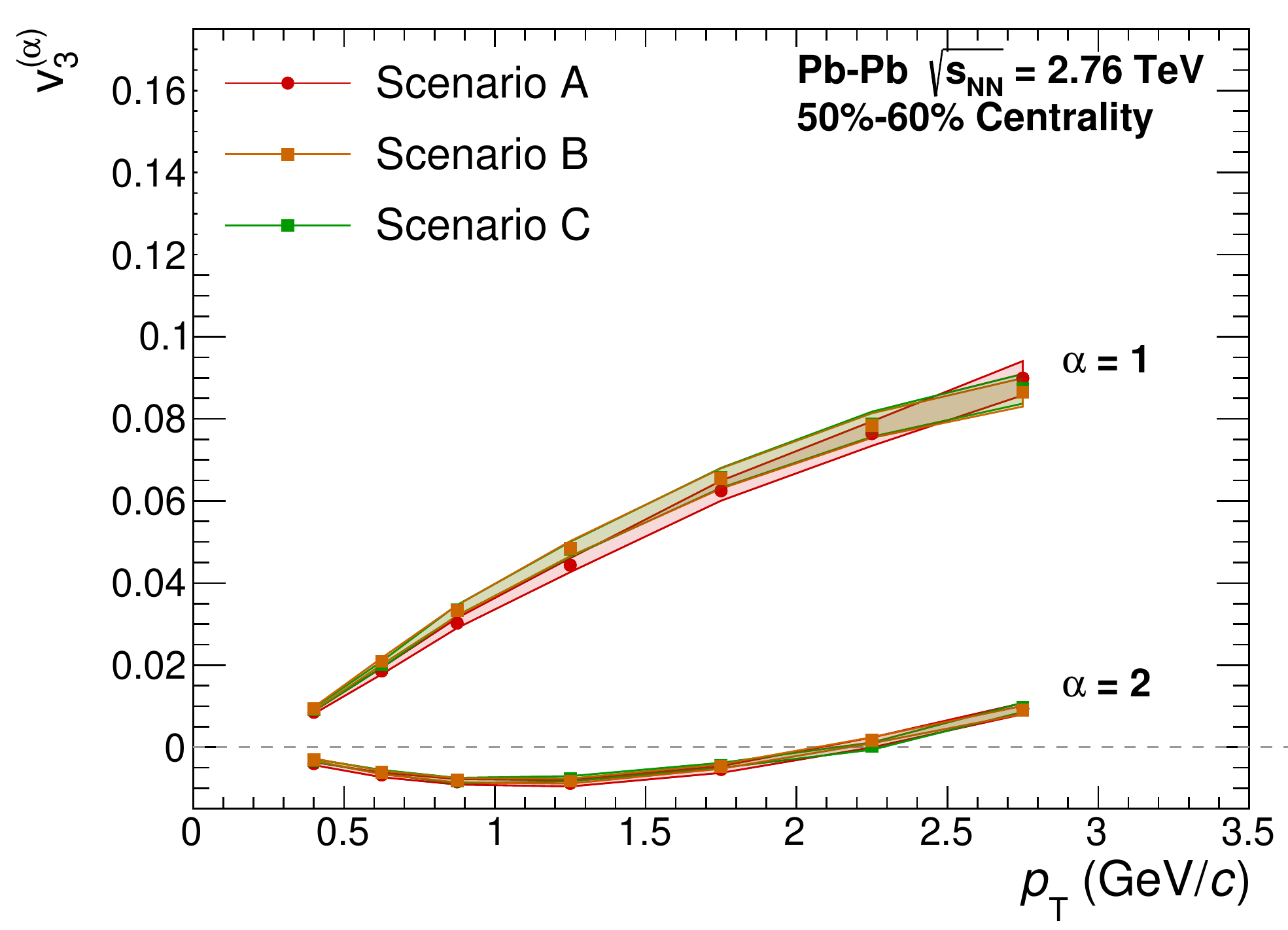}
  \caption{First two principal components of the two-particle correlation matrix according to the ExTrEMe prescription \cite{Hippert:2019swu}, plotted as a function of the transverse momentum $p_T$, for the harmonic $n=3$ and several centrality classes. }
  \label{fig:pca3app}
\end{figure}

\bibliography{ref}

\end{document}